# High-Performance Silicon Photonic Single-Sideband Modulators for Cold Atom Interferometry

**ASHOK KODIGALA[1*], MICHAEL GEHL[1], GREGORY W. HOTH[1†], JONGMIN LEE[1], CHRISTOPHER T. DEROSE[1], ANDREW POMERENE[1], CHRISTINA DALLO[1], DOUGLAS TROTTER[1], ANDREW L. STARBUCK[1], GRANT BIEDERMANN[1¥], PETER D. D. SCHWINDT[1], AND ANTHONY L. LENTINE[1]**

[1]Sandia National Laboratories, 1515 Eubank Blvd SE, Albuquerque, NM 87123, USA
†Present address: National Institute of Standards and Technology, 325 Broadway St, Boulder, CO 80305, USA
[¥]Present address: Homer L. Dodge Department of Physics and Astronomy and Center for Quantum Research and Technology, University of Oklahoma, Norman, OK 73019, USA
*Corresponding author. Email: akodiga@sandia.gov

**Abstract:**
The most complicated and challenging system within a light-pulse atom interferometer (LPAI) is the laser system, which controls the frequencies and intensities of multiple laser beams over time to configure quantum gravity and inertial sensors. The main function of an LPAI laser system is to perform cold-atom generation, state-preparation, state-selective detection, and to generate coherent two-photon process for the light-pulse sequence. Substantial miniaturization and ruggedization of the laser system can be achieved by bringing most key functions of the laser system onto photonic integrated circuit (PIC). We demonstrate a high-performance silicon-photonic suppressed-carrier single-sideband (SC-SSB) modulator at 1560 nm, which can dynamically frequency shift within the LPAI. With independent RF-channel control, we study the imbalances in both the optical and RF phases/amplitudes to reach 30dB carrier-suppression, unprecedented 47.8dB sideband-suppression at peak conversion-efficiency: -6.846dB (20.7%). Using a silicon photonic SSB-modulator, we demonstrate cold-atom generation, state-selective detection, and atom interferometer fringes to estimate gravitational acceleration, $g \approx 9.77 \pm 0.01$ m/s$^2$, in a Rubidium ($^{87}$Rb) atom system.

## INTRODUCTION

Light-pulse atom interferometers (LPAIs) have shown exceptional sensitivities to inertial forces in the laboratory making them perfectly suited for realizing high-performance inertial sensors [1-3]. However, current laser systems used for LPAIs are mostly based on discrete photonic components that are connected through fiber-to-fiber connections or free-space optical paths with optomechanical alignment mounts, which limit their ability to withstand high motional dynamics, limit manufacturing scalability, and make LPAIs less deployable [4-5]. At present, ensuring LPAI operation in high-dynamic conditions is hindered by fundamental challenges with miniaturization and ruggedization of an LPAI laser system [6-10]. As an alternative to a bulky laser system, photonic integrated circuit (PIC) laser architectures can be compact and robust to high-dynamics, e.g., a PIC-based laser system with integration of multiple PIC components on a single photonic chip. PIC technologies show great potential for a compact and high-performance laser system for quantum sensor applications (e.g., optical atomic clocks, atomic accelerometers, gyroscopes, gravimeters, gravity gradiometers, magnetometers, and electrometers) by reducing design complexity and improving reliability of the laser system [11-14]. Waveguide-to-waveguide connections between on-chip integrated photonic components (i.e., modulators, amplifiers, frequency doublers, and photo-detectors) guarantees robustness and reliability, which is necessary to realize extreme miniaturization of a laser system. We envision integrating all the required PIC components onto a single photonic platform to enable a highly miniaturized laser system [15-21] (Fig. 1). Here, we show a laser architecture based on hybrid integration with a key PIC component, a silicon photonic single-sideband (SSB) modulator, which can provide the required functions of manipulation of various optical frequencies and intensities time-multiplexed for LPAI operation [22-23]. This SSB modulator allows for the generation of multiple tunable coherent optical channels from a single laser (addressing the need for multiple independent lasers at various wavelengths) and the ability to offset-phase-lock two independent optical channels, which is an essential function for our LPAI architecture.

In recent years, integrated silicon photonics has matured rapidly to address an increasingly complex and broad application space. The main draw for silicon photonics is its ability to support the integration of complex photonic circuits at the chip-scale by taking advantage of standard complementary metal-oxide-semiconductor (CMOS) fabrication process which can be used for mass production with high yield. With growing demand for silicon photonics, there is a strong need for high performance GHz-scale optical single sideband (SSB) modulators/frequency-shifters on a silicon platform benefitting a variety of applications. These include frequency shifting or conversion for atomic physics research [22, 24-25], radio-over-fiber (RoF) communication systems [26-27], integrated microwave photonics [28-30], light detection and ranging (LiDAR) [31], high-resolution spectroscopy [32-33], and dense wavelength division multiplexed (D-WDM) networks [32]. However, to date, much of SSB work has been concentrated on a lithium niobate (LiNbO3) platform [34-36]. The few SSB modulators realized on silicon employing resonant ring modulators often lack carrier suppression and are not readily suitable for high-power applications [37-38]. Some realizations on a silicon platform make use of electro-optic polymers which have limited CMOS compatibility and suffer from high optical losses [39]. There are commercially available fiber or free-space acousto-optic modulators (AOMs) which can frequency-shift but are only limited to a few hundred MHz with GHz bandwidths available with significant loss in efficiency [40]. AOMs are also not easily integrated on-chip as they have an angular dependence of the output beam on the modulation frequency which significantly limits bandwidth [41]. Moreover, some of these SSB generation techniques require the filtering of one sideband from a double-sideband (DSB) output via a frequency matched notch or band-reject filter thereby limiting the bandwidth [32]. Another approach is the optical serrodyne modulation of an electro-optic modulator which can in principle achieve wide bandwidths with high efficiencies [42]. However, this is limited by the quality of the sawtooth waveform generated by an expensive arbitrary waveform generator (AWG) and becomes increasingly difficult for high gigahertz bandwidths [43]. A silicon photonic SSB modulator with dual-parallel Mach-Zehnder modulator (DP-MZM) configuration can provide a reliable and high-performance solution along with manufacturing scalability.

In this article, we design, fabricate, and characterize a silicon photonic suppressed-carrier single-sideband (SC-SSB) modulator using a dual-parallel Mach-Zehnder modulator (DP-MZM) configuration for atom interferometry. For SC-SSB generation, DP-MZMs have also been employed in silicon [44-45]. In their current form, these may not achieve sufficient carrier and sideband suppression (unwanted sidebands with respect to the wanted sideband) needed for atom interferometry [22,23]. In what follows, with independent RF control of amplitude and phase at each nested MZM, we optimize the carrier suppressed SSB generation to achieve higher performance than former studies. As a result, we simultaneously achieve an ultra-high sideband suppression of 47.8 dB (over the previous record, 39 dB [44]) and 30 dB carrier suppression at 1 GHz frequency shift. This result promises a high-performance PIC laser system for LPAI operation with reduced unwanted optical scattering and frequency shifts.

Our PIC approach will lead to reduced size, weight, and power (SWaP) of the quantum sensors, and, thus, represents an important step towards deployable sensors that are robust against vibration, shock, and radiation. This multidisciplinary union of integrated photonics and quantum physics reflects the current trend of significant research interest and investment being made in emerging quantum sensing technologies. In this paper, we study a PIC laser architecture which uses a telecom wavelength (1560 nm) for optical modulation and amplification along with a frequency-doubled wavelength (780 nm) to address rubidium ($^{87}$Rb) atoms. To validate this PIC laser architecture, we configured an exemplar laser system including a silicon photonic SC-SSB modulator chip that can provide all the required optical frequencies in a time-multiplexed manner (see Figure 1). With this laser system, we demonstrate laser-cooled atoms and state-selective detection for the atoms and realize a proof-of-concept atom interferometer. A single-channel SSB modulator was used to demonstrate state-selective detection. While a four-channel SSB modulator chip is depicted in Figure 1, only one SSB channel on this four-channel chip (fabricated) was used for the atom interferometer experiment. However, our approach is readily scalable to multiple SSB channels and an object of future study. Our on-going development efforts on each PIC component will enable a portfolio of modular PIC components for quantum sensing, which can be used for multiple high technology-readiness-level applications and increase market volume of PICs for quantum applications.

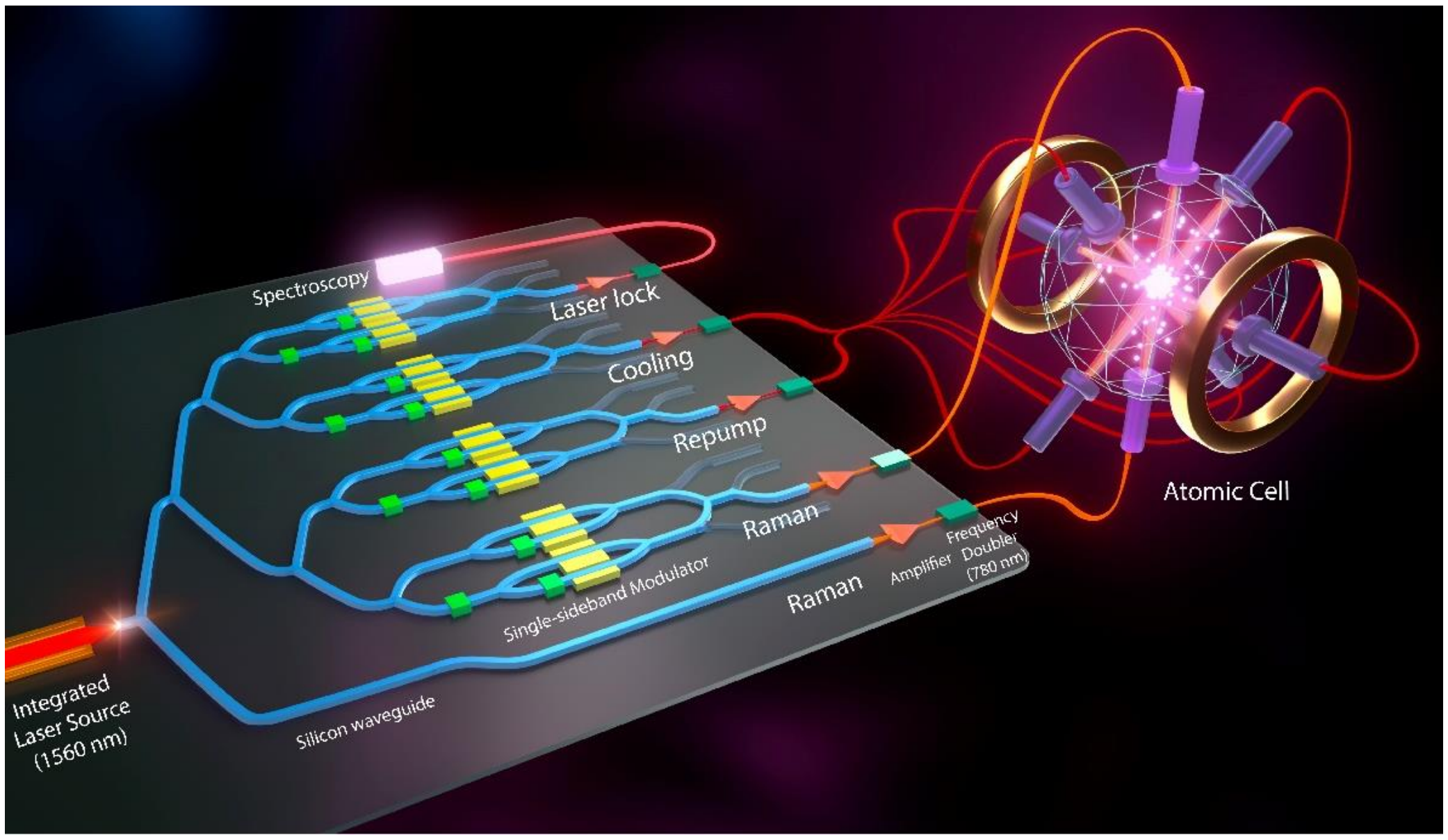

**Fig. 1. Conceptual rendering of a fully integrated multi-channel silicon photonic single-sideband (SSB) modulator chip generating tones supplied to a cold-atom interferometer experiment powered by a single integrated laser source.** Each silicon photonic modulator is based on a dual-parallel Mach-Zehnder modulator (DP-MZM) configuration. This DP-MZM is a key subcomponent of our photonic integrated circuit (PIC) laser architecture for quantum sensing technology. Optical amplifiers (triangle blocks) boost the power at the SSB modulator output. For each channel, a frequency doubling element (dark green blocks) converts sidebands near a telecom wavelength of 1560 nm to 780 nm thereby providing multiple light sources (laser lock, atom cooling, repump, detection, and Raman beams) for the cold-atom-based quantum sensing experiments. Atomic cloud (pink) is laser-cooled in vacuum by laser beams generated from the silicon photonic SSB modulator chip.

## 1. Integrated Photonics Architecture for Atom Interferometry

A laser system must be able to perform all functions required for LPAI operation, such as cold-atom generation, state preparation, state-selective detection, and matter-wave interference in order to support atom interferometry for quantum gravity and inertial sensing [22,23]. This LPAI laser system requires sophisticated frequency and intensity control of multiple optical channels over time, as well as frequency locking to an atomic transition and offset-phase locking between optical channels. As mentioned earlier, most existing LPAI laser systems in the laboratory currently have been configured using bulky optics sensitive to vibration, which include fiber-to-fiber or free-space optical connections with opto-mechanical mounts. Since these bulky laser systems cannot withstand high motional dynamics, the miniaturization and ruggedization of an LPAI laser system should address operation under such high dynamics [6-10]. Hence, we investigate a PIC based laser architecture to improve the robustness and reliability of an LPAI laser system with integrating multiple optical components into a single assembly.

We choose silicon photonic SSB modulators in our PIC laser architecture for atom interferometry to simplify generation of multiple, tunable coherent optical channels (originated from a single laser source) and shift the optical frequencies of each channel in a time-multiplexed manner for multiple LPAI operations. As described in [23], this PIC laser architecture is based on a 1560-to-780 nm frequency-doubling approach and hybrid integration, as shown in Fig. 2, which includes three major functional blocks: optical modulation (silicon photonics), optical amplification (III-V semiconductors), and optical frequency doubling (lithium niobate). A silicon PIC chip with ‘N’ SSB modulators can generate ‘N’ closely spaced and frequency-shifted optical channels near 1560 nm using a 1×N optical input splitter where ‘N’ is the number of optical channels. Within each channel, a SSB modulator creates an independent optical source by offset-locking the optical frequency for a particular function (*i.e.* Raman 1, Raman 2, Repump/Detection, Cooling/Depump, etc). A silicon PIC chip can provide all the needed optical channels using a single light source to realize cold atoms and operate atom interferometers. The chip would include thermo-optic phase shifters, variable optical attenuators (VOAs), optical filters, and photodetectors as power monitors. This approach allows for accurate frequency tuning and rapid switching of an optical channel in a time-multiplexed manner. The frequency accuracy and timing relationships are critical to the performance of an atom interferometer.

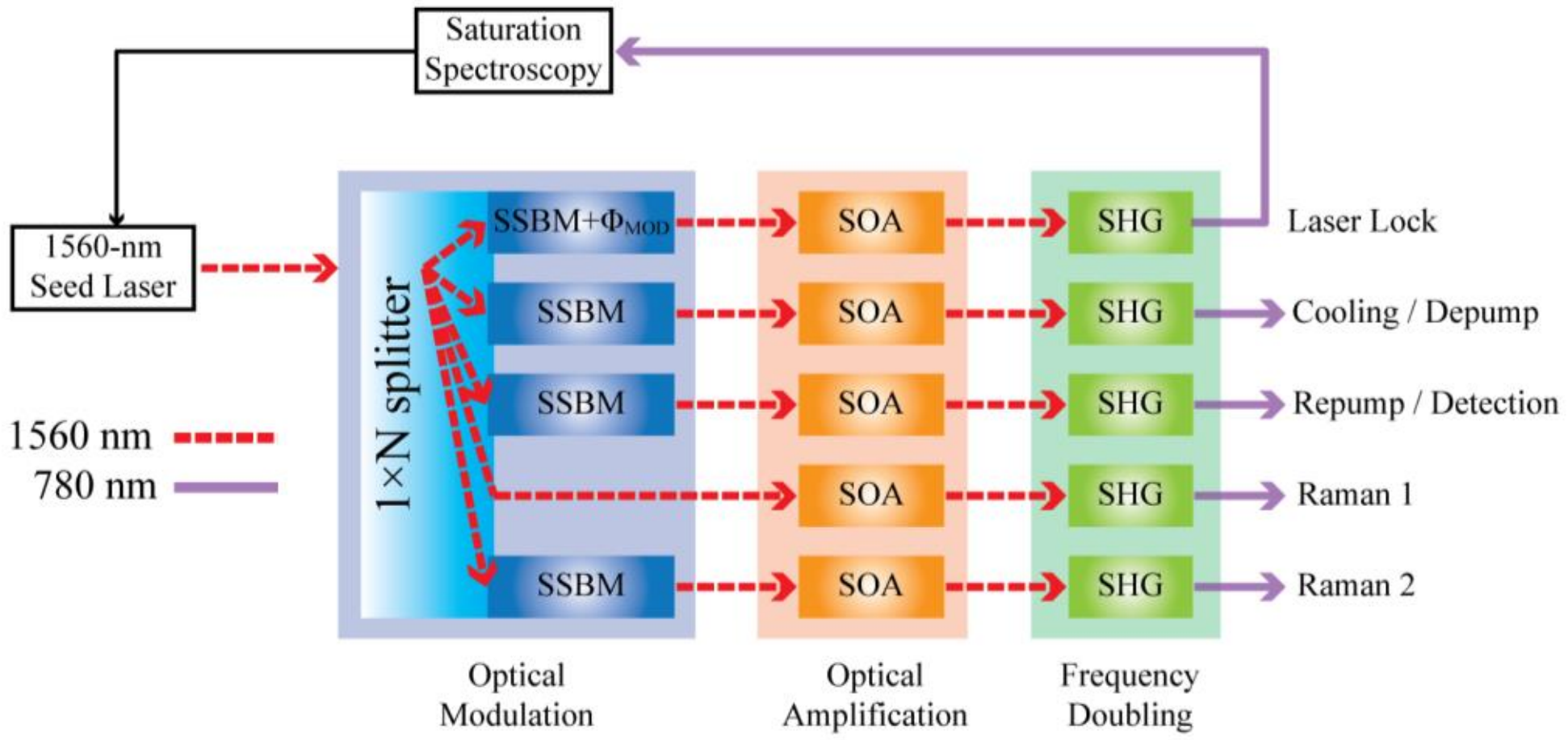

**Fig. 2. A PIC laser architecture based on 1560-to-780 nm frequency-doubling approach for atom interferometry.** A PIC-based laser system consists of three major functional blocks: optical modulation (silicon photonics), optical amplification (III-V semiconductors), and optical frequency doubling (lithium niobate) towards a single PIC platform via hybrid/heterogeneous integration and co-packaging, where SSBM is a silicon photonic SSB modulator, $\Phi_{MOD}$ is an additional on-chip phase modulator, SOA is a semiconductor optical amplifier, and SHG is a second harmonic generator [23]. Each channel in the silicon photonics includes a variable optical attenuator to control the optical amplitude.

The use of a silicon photonic SSB modulator in the PIC laser architecture for atom interferometry requires sufficient optical amplification at 1560 nm and efficient frequency doubling (from 1560 nm to 780 nm) to address rubidium atoms at 780 nm (see sections S1 and S2 for optical losses and power requirements). We chose this doubling approach primarily due to the maturity and availability of silicon photonics at telecommunications wavelengths (1560 nm). Moreover, there is greater availability of the needed low-noise and narrow linewidth laser sources at telecommunications wavelengths compared to 780 nm. However, a big drawback to this approach is the low frequency-doubling efficiency. Recent advances in heterogeneous integration (HI) of thin-film lithium niobate (TFLN) on silicon in realizing high-speed modulators at 1560 nm and separately periodically-poled TFLN for efficient frequency doubling make it feasible to fully integrate both LN modulators and doublers onto a silicon photonic chip [16,20]. This approach also allows for HI IIIV-on-Si laser sources (1560 nm) thereby providing a feasible option for integrating all components on the same integrated circuit minimizing coupling losses from one component to another (see Figure 1) [21]. However, it is challenging to integrate all components onto the same photonic chip due to fabrication and design complexity. Alternatively, recent demonstrations of TFLN modulators near 780 nm using partially etched LN or silicon nitride (SiN) waveguides make it possible to realize a multi-channel SSB modulator chip directly at 780 nm [46-48]. The drawback, however, is the integration of a 780 nm laser source or gain elements on the same chip which remains a challenge. Lastly, monolithic GaAs/AlGaAs PIC integration directly at 780 nm has the advantage of power efficiency and being able to integrate a laser, optical amplifiers, and modulators all on the same integrated circuit with the caveat of high optical waveguide losses.

Beyond hybrid PIC integration with co-packaging presented here, we are working to develop PIC laser subcomponents via heterogeneous and monolithic PIC integration for quantum applications towards the complete integration of nearly all of the many required optical components onto a single PIC chip [15, 16-21]. Additionally, for the atomic system itself, there are significant efforts to miniature atomic vapor cells with silicon/glass cells [49], compact vacuum packages [23], and passively-pumped long-term (>200 days) vacuum packages [50]. Many other custom electronic hardware component solutions need to be explored for complete miniaturization which include current sources, RF sources, RF hybrid couplers, AOMs, and feedback control electronics. All our efforts on PIC technologies and quantum applications will achieve reduced SWaP of quantum sensors, and, thus, make an important step towards deployable quantum gravity and inertial sensors robust to vibration, shock, and radiation.

# RESULTS

## 2. Background and Design

In what follows, we use a silicon photonic DP-MZM with independent RF control for achieving high-performance SC-SSB generation. A DP-MZM configuration cancels the unwanted sidebands and simultaneously suppresses the carrier thus alleviating the need for any output filters (see Fig. 3A) [34]. In a DP-MZM, the incoming optical signal is split into two arms with an optical phase delay applied on one. Each signal is sent through an intensity modulator or MZM that has its own optical phase control for carrier suppression and a pair of phase modulators that generate sidebands when driven by an RF signal. The outputs of each MZM are re-combined to form a single-sideband signal through interference. In a DP-MZM, the carrier suppression is primarily determined by the careful balancing of the optical amplitudes and phases, whereas sideband suppression is governed by the RF amplitudes and phases along with the optical extinction ratio in each nested-MZM (regions R2 and R3). In reference [44], the RF amplitude imbalance (between nested-MZM R2 and R3) is compensated with a counteracting optical power imbalance in each nested-MZM to achieve a high sideband suppression of 39 dB. However, this approach will not achieve peak SSB conversion efficiency as each nested-MZM is asymmetrically driven (*i.e.* RF amplitude imbalance) and also does not compensate for any RF phase imbalance [51]. The larger the RF amplitude imbalance, the worse the conversion efficiency at the high sideband suppression point with optical compensation. Previously, when imbalanced, we achieved a sideband suppression of 12 dB with the use of hybrid RF couplers [15]. With careful balancing of both the RF phases and amplitudes through independent RF control of R2 and R3 (different from the use of hybrid RF couplers [15]), we achieve high optical sideband suppression of 47.8 dB over the previous record of 39 dB [44].

As shown in Fig. 3A, our silicon photonic DP-MZM (with two nested MZMs as R2 and R3) has four carrier-depletion based electro-optic (EO) phase modulators (in yellow) along with several thermo-optic (TO) phase shifters (in green). The two nested MZMs are each optically balanced with equal optical path lengths. The two MZMs phases are optically (region R1) and electrically (R2 and R3) in quadrature ($\pi/2$). Each nested MZM has two EO phase modulators and a TO phase shifter. These two RF EO phase modulators are $\pi$ out of phase with each other and the TO phase shifter also has a $\pi$ optical phase shift.

The SC-SSB generation by a silicon photonic DP-MZM is as follows. First, carrier suppression at the output of the DP-MZM is achieved with the balancing of only optical phases ($\pi$ in R2 and R3) and amplitudes when RF power is turned off to all the EO phase modulators. Second, with the RF power turned on to the EO phase modulators, each nested MZM generates only odd harmonics with the carrier suppressed at its output. All the even harmonics cancel as the EO phase modulator are driven $\pi$ out of phase with each other. The relative optical phase shift of $\pi/2$ in R1 (green) along with the electrical $\pi/2$ phase difference between R2 and R3 (yellow) leads to a relative phase difference in the odd harmonics generated between the two Mach-Zehnder outputs. When combined, every other odd harmonic interferes constructively (+1,-3,+5,..) and others destructively at the output. The complementary harmonics (-1,+3,-5,..) can either be observed in an adjacent output port if present or are radiated out. We now have a carrier-suppressed single-sideband (-1) optical signal with intrinsically decaying unwanted higher-order sidebands (+3,-5,+7,..). This approach fundamentally limits the peak SSB conversion efficiency of the carrier into ±1 sidebands to -4.7 dB (~34%) [34]. Additionally, with perfect couplers and phase-shifters, there are no even harmonics. However, in reality, this is not the case and even harmonics are also present due to non-linearities in the carrier based EO phase modulators.

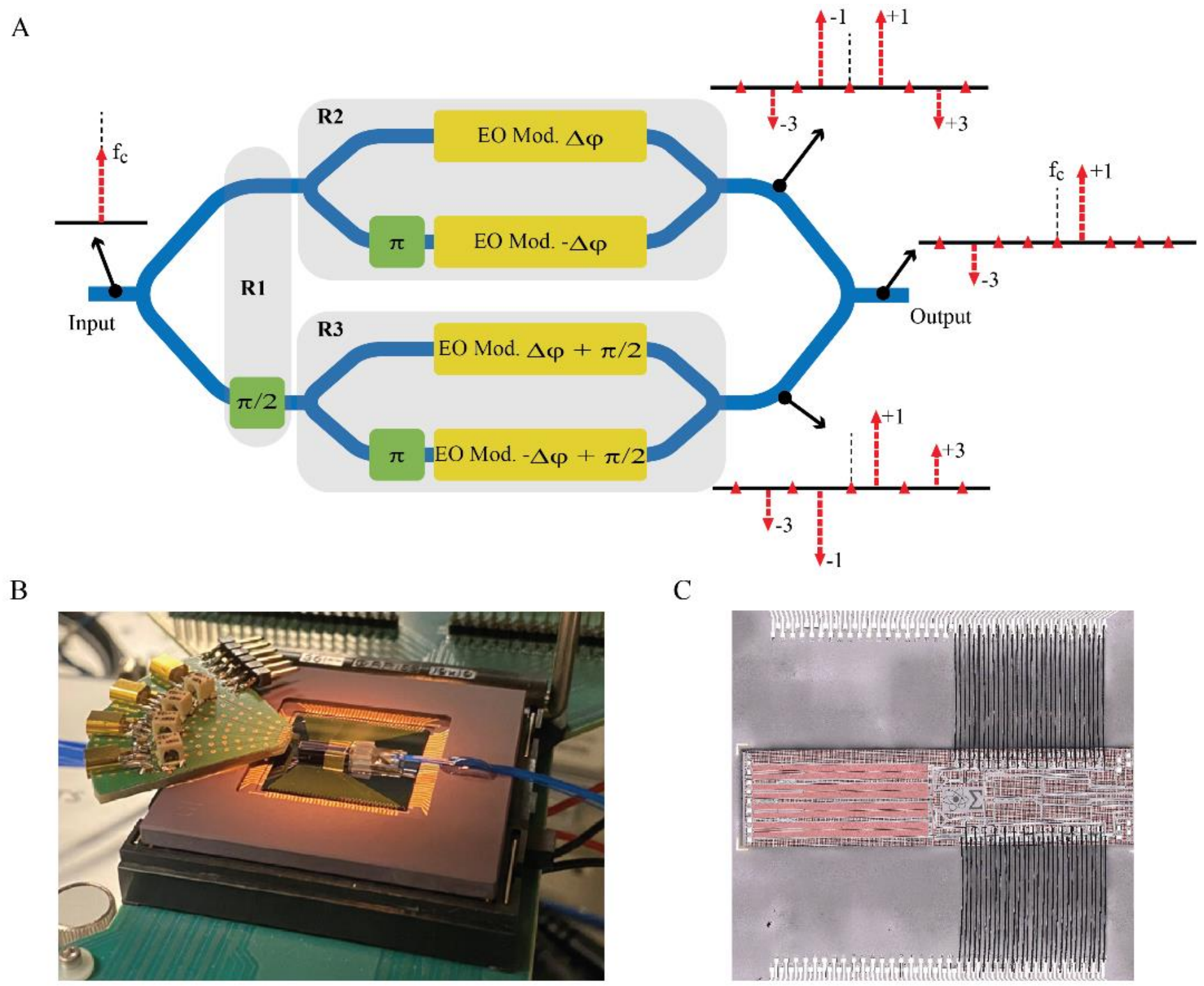

**Fig. 3. Sandia-developed silicon photonic modulator for suppressed-carrier single-sideband (SC-SSB) modulation.** (**A**) Operational schematic of a DP-MZM which has three regions (R1, R2, and R3) with regions R2 and R3 in parallel and each consisting of an MZM. R1 has a thermo-optic (TO) phase shifter (Left, Green), and R2 (Top nested-MZM) and R3 (Bottom nested-MZM) have a TO phase shifter each (Green) and two electro-optic (EO) phase-modulators (Yellow) with appropriate RF phase offsets, respectively. Independent control of R2 and R3 enables high-performance SC-SSB generation with our modulator. (**B**) Picture of an optically and electrically packaged one-channel silicon-photonic single-sideband modulator showing a v-groove array for optical lines and wire bonds for DC and RF to PCB. (**C**) Top-view of a fabricated silicon photonic DP-MZM modulator with DC wire bonds to an interposer chip with each nested MZM in a push-pull configuration making up the DP-MZM.

Our silicon photonic SSB modulators are fabricated at Sandia's Microsystems Engineering, Science and Applications (MESA) complex (see Materials and Methods). As Fig. 3B shows, our SSB modulator is packaged both optically and electrically, with a fiber v-groove array, wire bonds to the interposer for DC biases (TO phase shifters), and wire bonds to the printed circuit board (PCB) board for RF signal lines. As each nested MZM is configured to be driven in a push-pull configuration with a single RF input, only two RF inputs are required for the full SC-SSB modulator. The carrier-depletion based EO phase modulators in silicon with doped regions are 1.55 mm in length (Fig. 3C). The chip has multi-mode interference (MMI) couplers for optical splitters and combiners. The outer Mach-Zehnder of a DP-MZM and the nested MZMs (R2 and R3) are optically balanced (same path lengths), therefore there is no wavelength dependence.

## 3. Experimental Results

We implement a self-heterodyne measurement setup to validate the performance of SC-SSB generation using our silicon photonic SSB modulators, as shown in Fig. 4. The measurement setup consists of a CW telecom-wavelength input laser (Keysight 81606A), polarization controllers, an acousto-optic modulator (AOM), fiber couplers/splitters, and a photodiode. The optical input is split into two paths, one through our SSB modulator chip and the other through a fiber coupled AOM and re-combined thereafter. The optical beat-note signal from combining two separate optical paths (one from the silicon photonic modulator output and the other from a frequency shifted AOM output) is detected by a photodiode and analyzed by an RF spectrum analyzer (Agilent 86146B). This approach ensures the optical resolution and signal integrity required to view all sidebands in the RF domain, which would be difficult with the commonly used scanning Fabry-Perot interferometer [15]. We test our SSB modulator with an independently controlled two-channel RF source (Holzworth 9004B) with signals RF Ch1 and RF Ch2 to drive nested MZM regions (R2 and R3). This allows us to independently control the amplitude and phase of each RF signal. Each nested MZM is driven in a push-pull configuration with a single RF input channel. We use an RF amplifier for each RF channel to reach peak SSB conversion efficiency (not pictured). Lastly, there are three current sources tuning the three TO phase shifters (DC1 to $\pi/2$, and DC2 and 3 to $\pi$) for optimal optical performance.

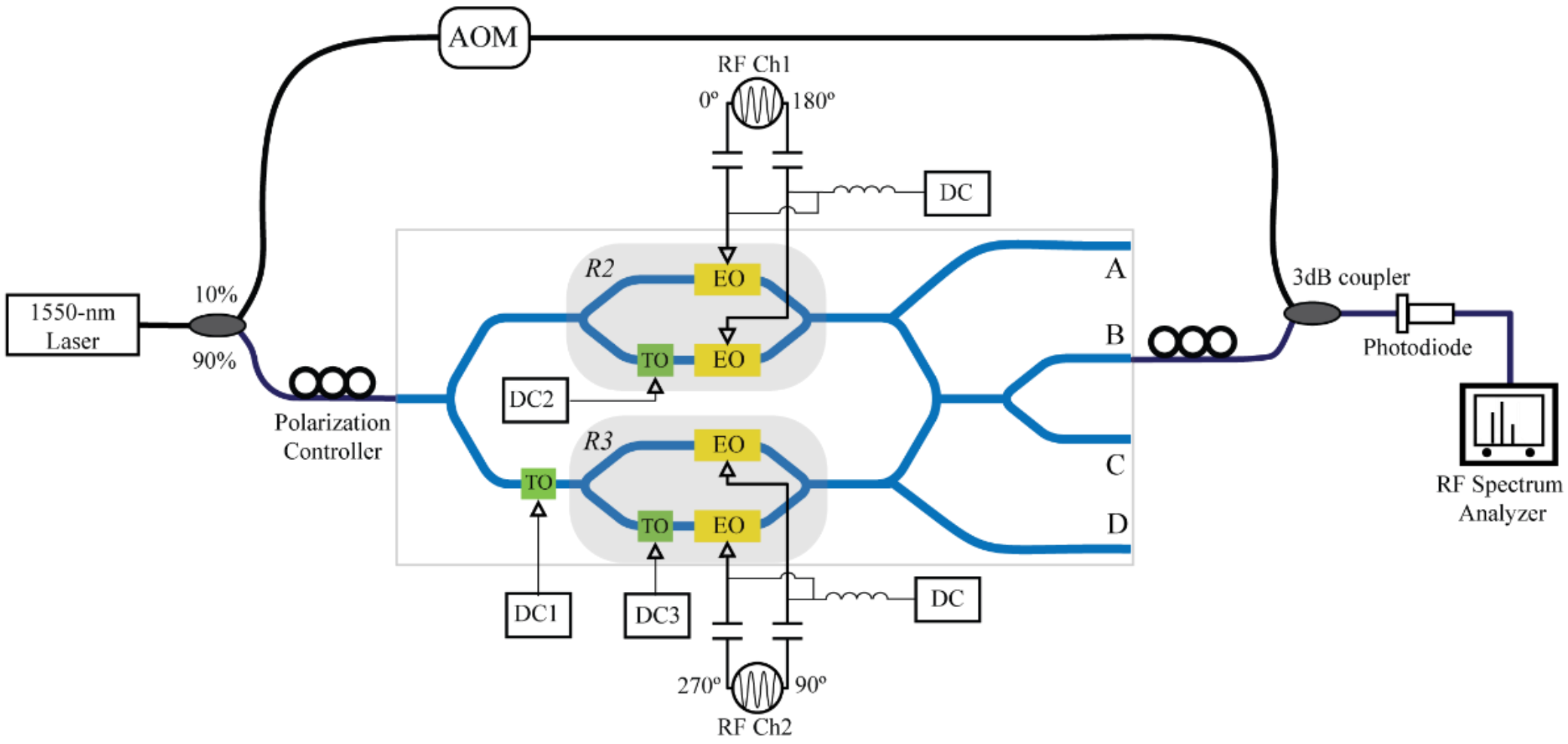

**Fig. 4. Experimental setup for characterizing our SC-SSB modulator chip.** We characterize the SSB modulation with a self-heterodyne measurement setup consisting of a CW telecom-wavelength input laser (Keysight 81606A), polarization controllers, an acousto-optic modulator (AOM) with frequency of 110 MHz, fiber couplers/splitters, and a photodiode. The optical beat-note signal from combining two separate optical paths (one from the silicon photonic modulator output and the other from a frequency shifted AOM output) is detected by a photodiode and analyzed by an RF spectrum analyzer. Our DP-MZM has four output ports (A, B, C and D). In full operation, the two inside output ports (B and C) output the sideband signals whereas the two outside output ports output the unwanted optical carrier. For RF inputs, we independently drive regions R2 and R3 (highlighted in gray) with two separate RF channels (RF Ch1 and RF Ch2) from a single multi-channel RF source (Holzworth 9004B). Each nested-MZM (R2 or R3) is designed to operate in a push-pull configuration. All DC and RF control lines are shown as black arrows for both TO phase shifters and EO modulators.

First, we calibrate for the optical losses of our modulator device including on and off chip fiber coupling losses (see sections S1 and S2 for optical losses and power requirements). With the RF turned off (no EO modulation), we first set our main DC biases (DC2 and 3 to $2\pi$) to maximize the optical carrier in our side-band ports (ports B and C) then set our remaining DC bias (DC1 to $\pi$ or $2\pi$) thereby directing all carrier signal to a specific sideband port (port B pictured). The optical power measured here captures all the optical chip-coupling and propagation losses in our SSB modulator and serves as a normalization for all subsequent sideband measurements. As a second step of the calibration, we then proceed to set our biases (DC2 and 3 to $\pi$) to suppress the carrier in the sideband ports re-directing all the carrier to the outermost ports (ports A and D). This suppresses the carrier by >40 dB, with RF still off. This suppression is limited by the extinction of each nested MZI (~30-40 dB). This can be further improved to >60 dB, with the use of high-contrast splitters [51, 52-55]. For the remainder of the testing, each carrier-based EO phase modulator had no reverse-bias due to a fabrication mask error preventing electrode access. Reverse biasing prevents the modulators from entering the forward bias region under a strong RF drive. Forward biasing the modulators drastically increases carrier concentration in the waveguide region and increases free-carrier absorption of the optical signal. With reverse-biasing, we expect lower optical losses and better linearity of our phase modulators and consequently improved SSB performance.

With the carrier now suppressed as described above, we turn on the RF signals to the EO phase modulators and set both the quadrature phases (one optical and one RF). We optimize together the optical phase (DC1 to $\pi/2$) and the relative RF phase ($\pi/2$) of RF-channel 1 (nested MZM in R2) to RF-channel 2 (nested MZM in R3) in order to maximize the extinction between +/-1 sidebands. Extinction between +/-1 sidebands is maximum at quadrature. Hence, by observing the +/-1 sidebands at the RF spectrum analyzer, a quick optimization scanning routine is run to adjust the bias current, DC1, setting the optical phase to quadrature. The same is done to adjust the RF-channel 1 phase to quadrature using the settings of the two-channel RF source. All the while, we ensure RF-amplitudes of both channels are equal. We now have our suppressed carrier single-sideband signal at port B and its complementary sidebands at port C.

### 3.1 Single-Sideband Conversion Efficiency

To reach peak SSB conversion efficiency, we then sweep the input RF amplitudes (powers) in each channel together while subtly optimizing (automated) DC1 and RF-amplitude of one channel to account for any minor electrical and optical imbalances over time with the sweep. The sweep was only performed once; however, the optimization was performed at each input RF power level. As shown in Fig. 5A, the resulting sidebands have Bessel function amplitudes and scale to the order of the sideband with modulation power. Hence, a first order sideband has a slope of 1 in log-scale, a second order sideband has a slope of 2 and so forth. At low RF powers ($P_{RF}$ per channel < +8dBm), we simultaneously observe a remarkable carrier suppression of >30 dB with unwanted sideband suppression of ~50dB.

We achieve peak SSB (-1) conversion efficiency of -6.846 dB (20.7%) at +21 dBm RF-power corresponding to a $V_\pi$ of ~2.5 V in each modulator arm ($V_\pi \cdot L$ = 0.388 V·cm). This difference from the efficiency limit of -4.7 dB can be attributed to the lack of reverse-bias in the modulators and inherent non-linearities of the silicon phase modulators [56-57]. These non-linearities also contribute to a rise in all other unwanted sidebands irrespective of RF and optical balancing at high RF powers. In this particular sweep, the automated optimization briefly fails at high RF powers ($P_{RF}$ per channel >19 dBm) and the high sideband suppression is lost shown by the dotted green line (+1). This is also likely from an increased non-linearity in the carrier-based phase modulators at high RF powers and needs to be further investigated. The linearity of the modulator can be improved with reverse-biasing.

Moreover, the carrier (0) suppression degrades with increasing RF-power indicating the non-linear nature of the modulators. In the absence of non-linearities, the carrier suppression is expected to be constant with RF signal (*i.e.* independent of RF power and phase) while only dependent on the optical amplitude and phases at each nested MZM. Near peak conversion efficiency at +19 dBm RF-power, we achieve a carrier suppression of 30 dB and extremely high sideband suppression of 47.8 dB (~50 dB) as shown in Fig. 5B. Our results nearly match the sideband suppression ratio achieved in a Brillouin laser system (49 dB) [58]. However, our silicon photonic SSB modulator is not limited to a MHz tuning bandwidth and does not require an external pump laser (see section S3 for details on modulator bandwidth). In future work, we expect to exceed the carrier and sideband suppression achieved here by implementing reverse-bias, improving both RF and optical amplitude and phase balancing.

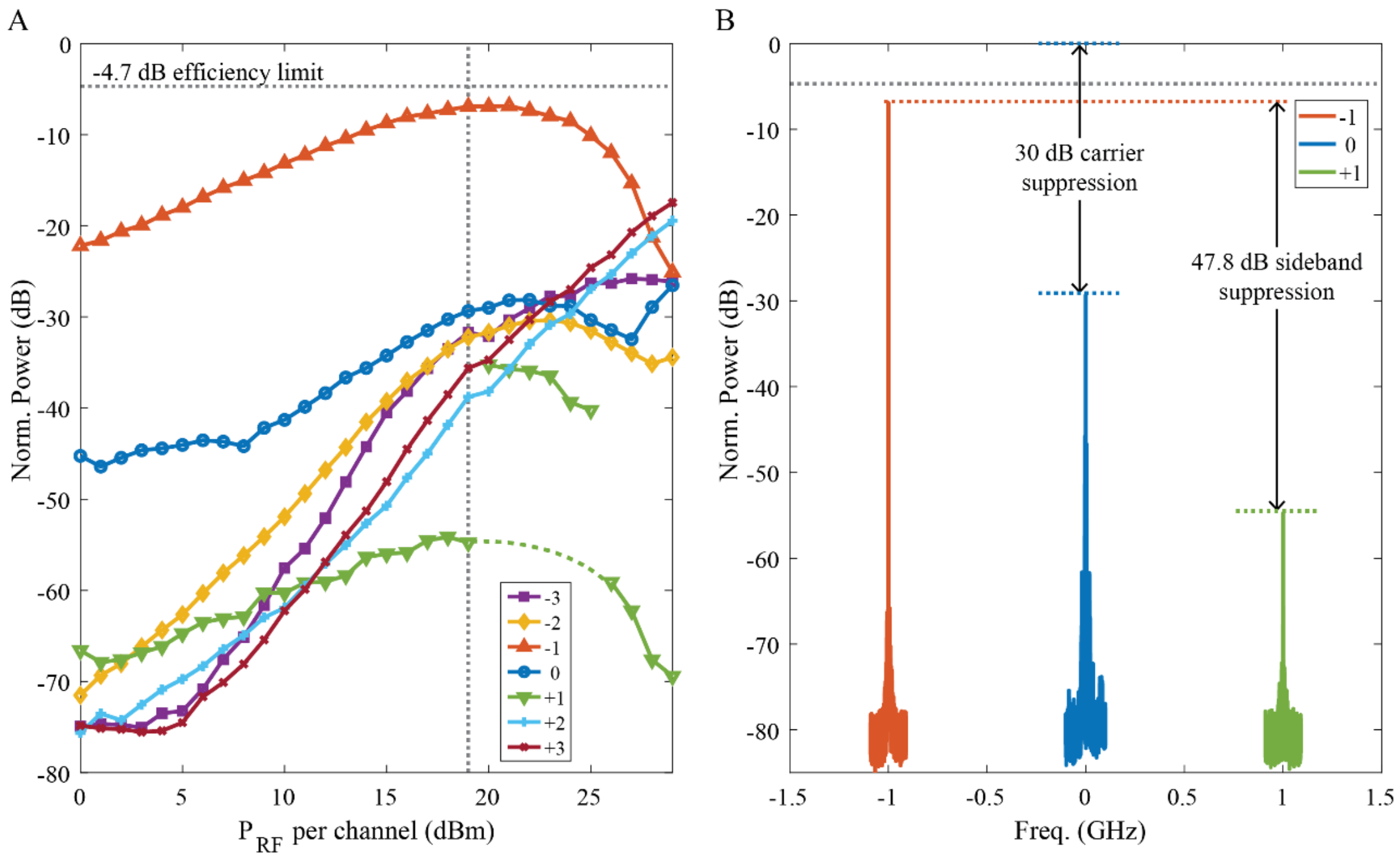

**Fig. 5. Measurement of high-performance SC-SSB generation.** (**A**) Measured optical powers of suppressed-carrier (0) and sidebands (±1, ±2, ±3) as a function of RF drive power in each channel. All powers normalized to the total on-chip optical carrier accounting for optical fiber coupling losses. Peak conversion efficiency of -6.84 6dB (20.7%) is achieved at +21 dBm RF-power corresponding to a $V_\pi$ of ~2.5 V in each modulator arm ($V_\pi L$ = 0.388 V·cm). At higher RF powers, non-linearities in the silicon modulators contribute to a rise in all unwanted sidebands irrespective of RF balancing. (**B**) Spectra of carrier and ±1 sidebands with a carrier suppression of 30 dB and sideband suppression of 47.8 dB at +19 dBm RF-power (vertical dotted line in left plot) at 1 GHz frequency.

## 3.2 RF Amplitude Imbalance

To better understand the effect of RF amplitude imbalance on SC-SSB performance, we generate a RF amplitude imbalance between the two RF channels ($\Delta P_{RF}$ from -1 to +1 dBm) with RF phases ($\Delta\phi_{RF}$= constant) balanced at quadrature (see Fig. 6). Here, we again optimize the modulator for sideband extinction at an RF power of +10 dBm in each channel with RF phases at quadrature. RF imbalances (both amplitude and phase) are common for commercial RF hybrid couplers and are known to limit device performance. We numerically model the response of an ideal silicon photonic SSB modulator to an RF amplitude (power) imbalance between the two nested MZMs (see Fig. 6A). As expected, when the carrier is fully suppressed, only the ±1 and +3 sidebands are present. According to the model, we observe that even a minor imbalance in RF powers results in severe degradation in sideband extinction. Achieving a sideband suppression greater than 40 dB requires a very limited RF amplitude imbalance of ±0.2 dB. Experimentally, we observe this predicted degradation in sideband suppression with amplitude imbalance and observe the tight tolerance required for high sideband extinction (>40 dB) (see Fig. 6B). We also note that the sideband suppression is nearly ~50 dB when RF amplitudes of the two channels are perfectly balanced ($\Delta P_{RF}$=0). Moreover, higher-order sidebands (±2, -3) are also present due to non-linearities in our modulator.

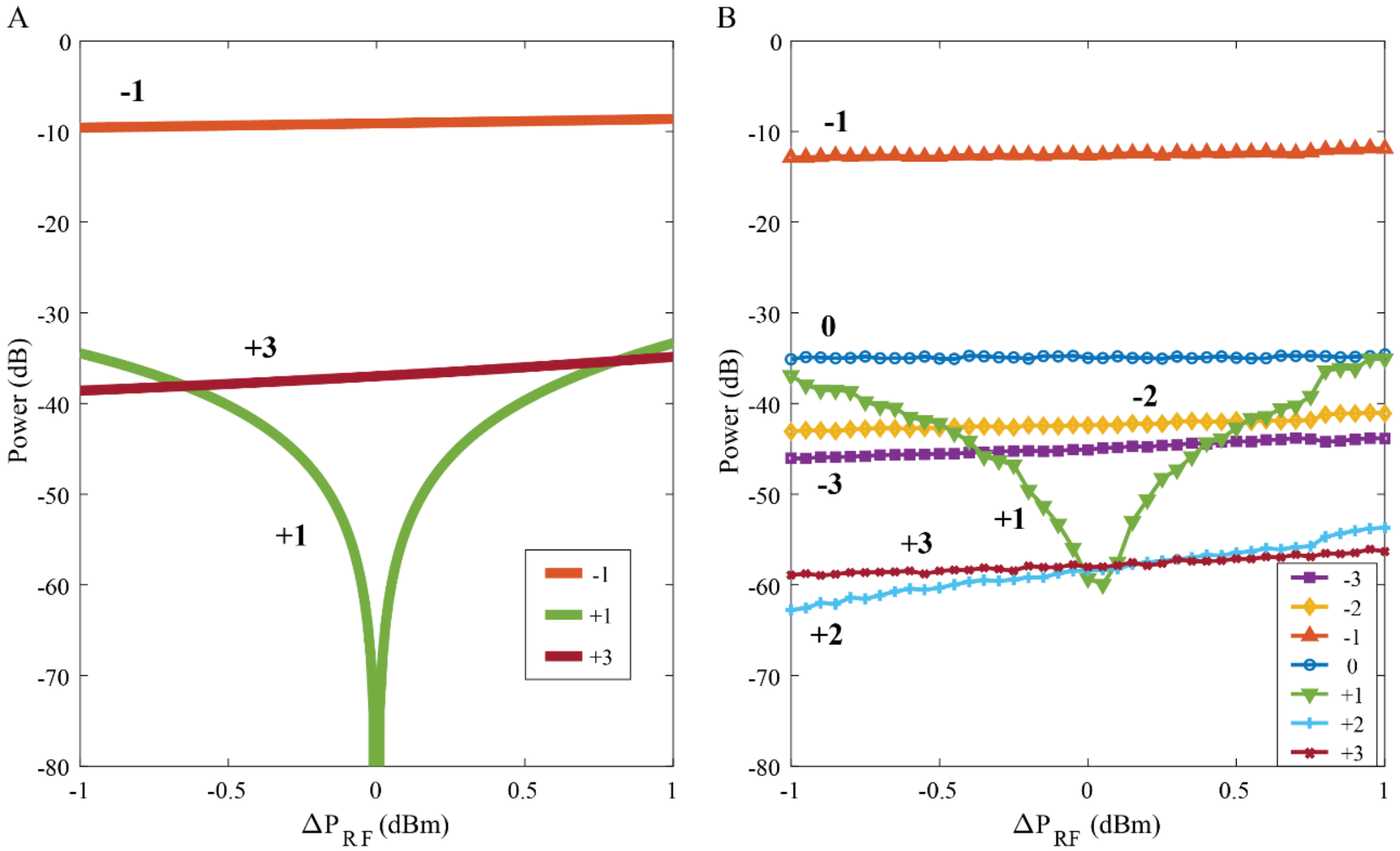

**Fig. 6. The effect of unbalanced RF amplitudes on SC-SSB performance.** (**A**) Model of an ideal SSB modulator operating with an RF amplitude (power) imbalance ($\Delta P_{RF}$ from -1 to +1 dBm) between RF channels (or nested MZMs) at a starting RF power in each channel of +10 dBm and RF phases at quadrature. In the ideal model, only the ±1 and +3 sidebands are present, as the carrier and other sidebands are completely suppressed. The model predicts that to achieve a sideband suppression > 40 dB, the RF amplitude imbalance must be within ±0.2 dBm. (**B**) Measured optical powers of suppressed carrier (0) and sidebands (±1, ±2, ±3) as a function of RF amplitude imbalance. For both simulation and measurement, we observe a severe sensitivity to amplitude imbalance on sideband suppression. We experimentally achieve a sideband suppression of ~50 dB when perfectly balanced ($\Delta P_{RF}$ = 0).

### 3.3 RF Phase Imbalance

Next, to study the effect of RF phase imbalance on SC-SSB performance (see Fig. 7), we introduce a phase imbalance ($\Delta\phi_{RF}$ from 0˚ to 360˚) from quadrature with balanced RF amplitudes ($\Delta P_{RF}$=0). Here, $\Delta\phi_{RF}$ of 0˚ indicates the two RF channel phases are at quadrature (π/2 or 90˚). Again, we optimize all the optical phases for carrier suppression and maximum sideband suppression. Our model and experimental measurements both indicate a tight phase tolerance (±5˚) for achieving sideband suppression greater than 30 dB (see Fig. 7A and 7B).

We also note that, as expected, the ±1 sidebands switch every 180˚ from quadrature in our model and the +3 sideband is periodic every 120˚. Experimentally, however, this switch for ±1 happens at ~230˚ instead which again could be attributed to the non-linearities in the modulator and needs further investigation (see Fig 7B). We also observe the periodicity of 120˚ in both +3 and -3 sidebands present. Higher order ±2 sidebands are once again present due to non-linearities in our modulator. Nevertheless, there is good agreement between the model and experiment for RF phase balancing.

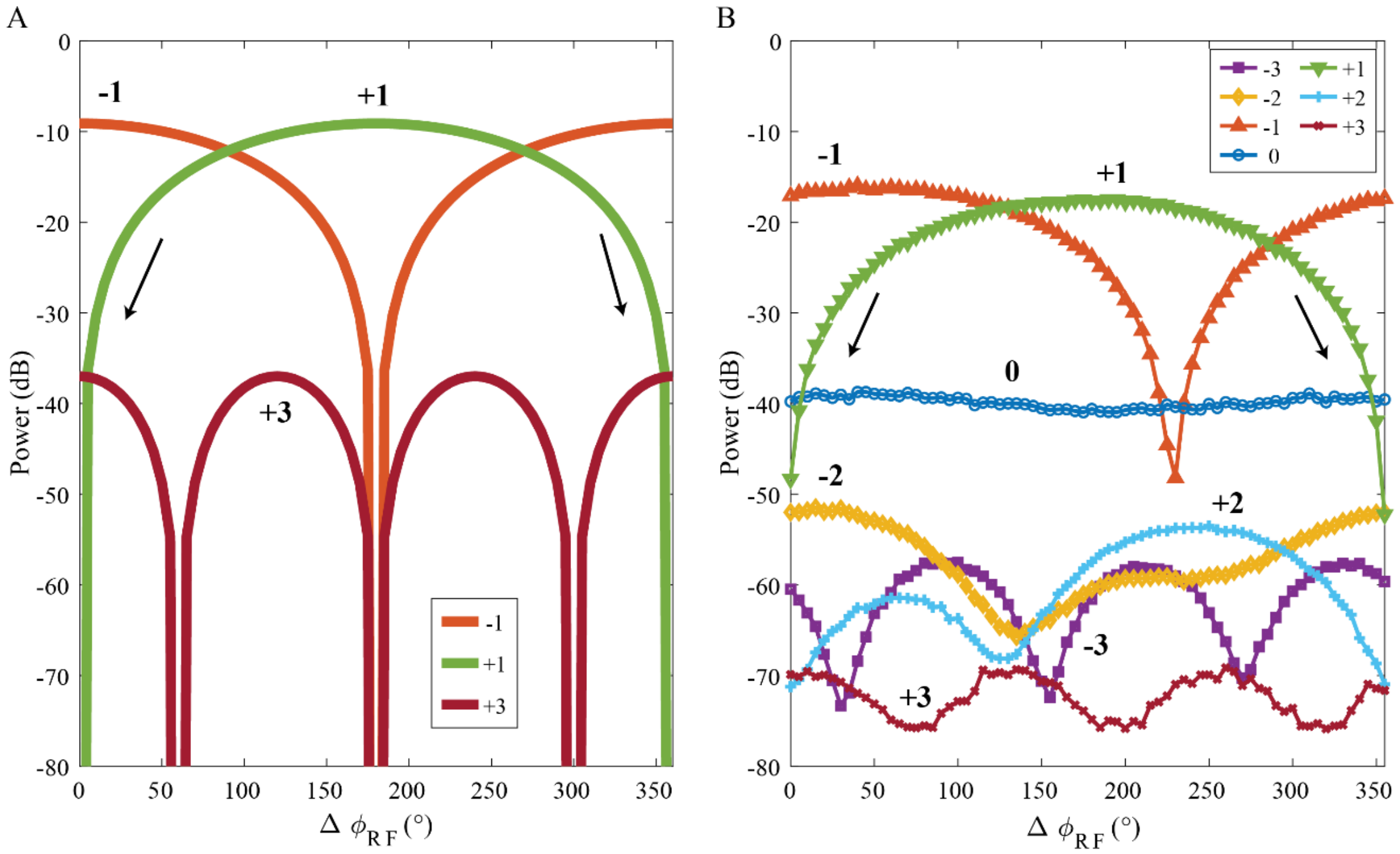

**Fig. 7. The effect of unbalanced RF phases on SC-SSB performance.** (**A**) Model of an ideal SSB modulator operating with a RF phase imbalance ($\Delta\phi_{RF}$ from 0° to 360°) between the two RF channels (or nested MZMs). The RF power to RF Ch1 and RF Ch2 is set to +6 dBm and RF phases at quadrature (90°). As expected, the ±1 sidebands switch at 180° with only ±5° tolerance for sideband suppression >30 dB. (**B**) Measured optical powers of suppressed carrier (0) and all sidebands (±1, ±2, ±3) as a function of RF phase imbalance ($\Delta\phi_{RF}$ from 0° to 360°). There is overall very good agreement with the model, however the phase at which the sidebands switch is at ~230° instead of 180° which needs further investigation.

We can further improve the peak efficiency of the SSB modulator by reverse-biasing the modulators which helps with linearity. In terms of optical balancing, the use of high-contrast splitters or cascaded MZMs will drastically improve the present carrier-suppression by increasing the extinction of each MZM. Along with the carrier, this also improves suppression of all even-harmonics (±2 sidebands). Moreover, the optical phases can be self-biasing with the implementation of an active feedback scheme operating in conjunction with integrated germanium (Ge) detectors as integrated power monitors [59]. As for RF balancing, we have assumed here that a pair of phase-modulators in each MZM are driven perfectly out-of-phase (π) in a push-pull configuration. In reality, this is not the case due to the inherent non-linearity of these carrier-depletion phase modulators with bias. Hence, a refractive index change in the positive portion of the sinusoidal RF drive is not perfectly equivalent to the index change in negative portion for a phase modulator. Our modelling indicates that a deviation from π will adversely affect even-harmonics (±2 sidebands suppression) including the carrier-suppression. This could be overcome with the use of a single-drive configuration where each modulator is independently driven but this comes at the cost of added complexity of having four RF channels to balance instead of the two presented here and is an object of future study.

## 4. Cold Atom Interferometry with Silicon Photonics

We envision a PIC-based laser system where all required PIC components are hybrid-integrated into a single photonic platform for optical modulation, amplification, and doubling. However, we first focus on validating a silicon photonic SSB modulator in an LPAI laser system. With our modulator we demonstrate cold-atom generation, state-selective detection for atoms with time-multiplexed frequency shifting, and demonstrate atom interferometer fringes to estimate gravitational acceleration (see Table 1 for tones generated).

**Table 1: Frequency shifts generated by silicon photonic SSB modulator chips and their purpose.**

| Wavelength | SSBM Characterization<br><br>(Fig. 5, 6, 7) | State-Selective Detection Demo<br><br>(Fig. 8) | Atom Interferometer Demo<br><br>(Fig. 9) |
|---|---|---|---|
| 1560 nm | ±1 GHz at 1550nm | +1.644 GHz to<br>-1.644 GHz switching | +3.417 GHz |
| 780 nm | N/A | +3.288 GHz (Repump)<br>-3.288 GHz<br>(Detection/cooling)<br>$f_{HF}$=6.576 GHz | $f_{HF}$=+6.834 GHz (Raman) |

To validate cold-atom generation and state-selective detection with rubidium atoms, we implemented a laser system including our single-channel silicon photonic SSB modulator (Fig. 8A-B). As previously described, this laser system starts from a single 1560-nm seed laser and has three major components: optical modulation with a silicon photonic SSB modulator, optical amplification using an erbium-doped fiber amplifier (EDFA), and frequency doubling with a second harmonic generator (SHG). The 1560 nm fiber laser with narrow-linewidth (≤3kHz, NP Photonics RFLM) is frequency-doubled and locked to an atomic transition using saturation spectroscopy. The silicon photonic SSB modulator produces repump and detection (or cooling) beams in a simultaneous or time-multiplexed manner at 780 nm for the cold-atom system to generate a magneto-optical trap (MOT). The SSB modulator is driven at 1.644 GHz at 1560 nm (3.288 GHz when doubled to 780 nm). As shown in Fig. 8C, a MOT was implemented in a compact ultra-high vacuum chamber (volume ~ 67 $cm^3$) using a six-beam optical geometry. After cold atom generation with the MOT, the hyperfine-ground-state atoms in $F$=1 or $F$=2 were prepared using an optical pumping step between the MOT phase and the detection phase.

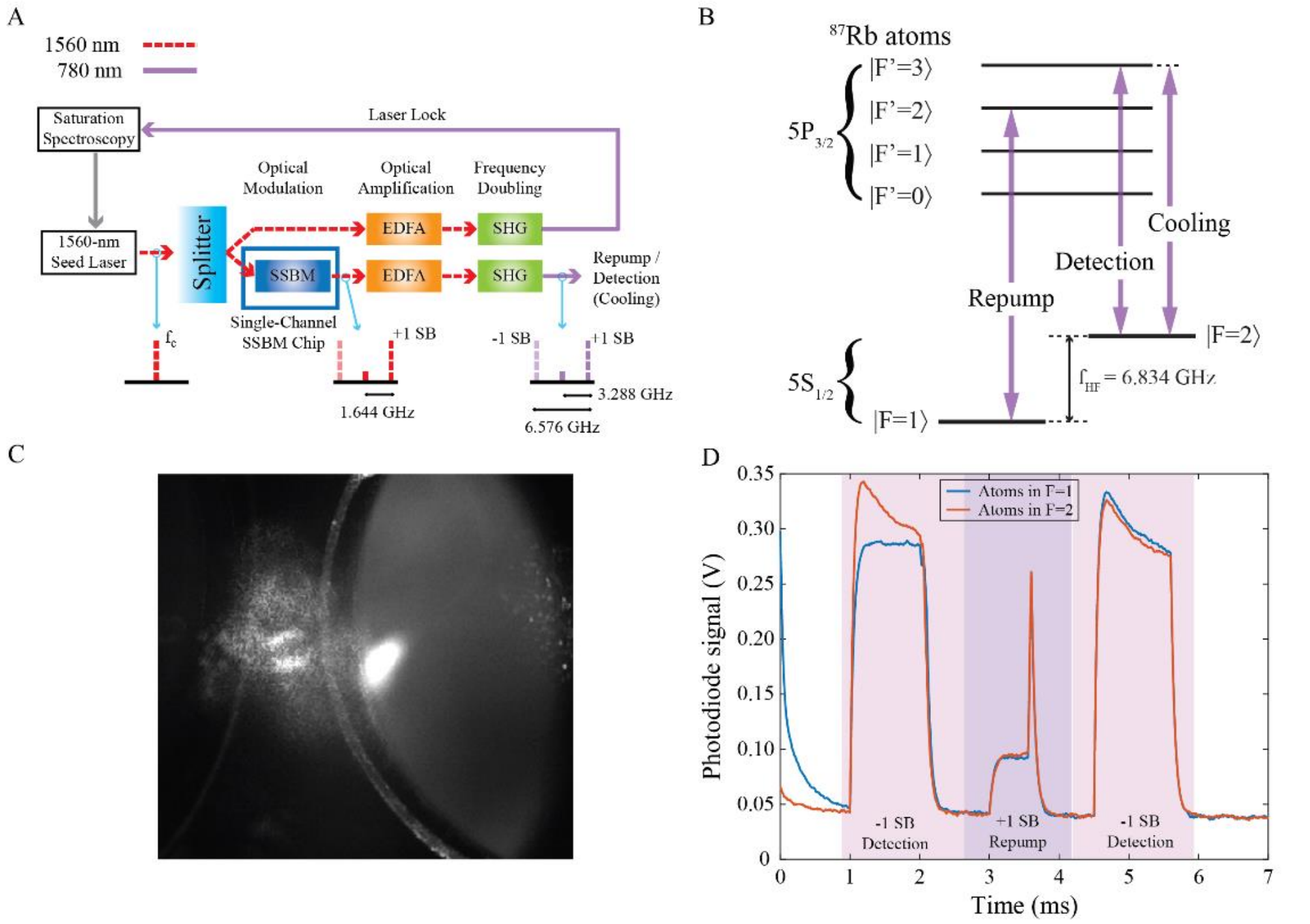

**Fig. 8. Cold-atom experiment with the use of a single-channel silicon photonic SC-SSB modulator in an LPAI laser system.** (**A**) Experimental setup for cold-atom generation and state-selective detection with rubidium atoms (780 nm), where SSBM is a silicon photonic SSB modulator, EDFA is an erbium-doped fiber amplifier, SHG is a second harmonic generator, and ±1 SB is ±1 sideband. (**B**) Atomic transitions ($^{87}$Rb D2 transition) related to cooling/repump/detection beams. (**C**) Picture of cold atoms in a magneto-optic trap (MOT) achieved via the simultaneous generation of cooling and repump beams. (**D**) Demonstration of state-selective detection for atoms in the initial state of $F$=1 (blue) or $F$=2 (red). On-chip TO phase shifter (response time = ~20 μs) is used to switch between detection (-1) and repump (+1) sidebands. The master laser is locked at the midpoint of detection and repump frequencies, so the total frequency jump at 780 nm is 6.576 GHz (3.288 GHz at 1560 nm). The sideband (or bias) switching corresponds to switching the quadrature optical phase between π/2 and 3π/2, and the intensity spike at the end of the second pulse of repump beam is caused by the sudden change in the bias current.

For atoms in the initial state of the $F$=1 or $F$=2, we demonstrate state-selective detection by sequentially generating detection, repump, and detection beams as shown in Fig. 8D. Here, the 1560-nm laser is locked between the two $^{87}$Rb transitions. An on-chip TO phase shifter is used to switch the output optical frequency between the -1 sideband (detection at -3.288 GHz for the $F$=2 to $F$'=3 transition) and the +1 sideband (repump at +3.288 GHz for the $F$=1 to $F$'=2 transition) with the total frequency jump at 780 nm being 6.576 GHz. The response time for our TO phase shifters are ~20 μSec. This state-selective detection process is needed for normalized atomic population measurement, the pulse sequence of which is detection, repump and detection beams.

In the first pulse of detection beam (-1 sideband), the atoms are illuminated with light resonant with the $F$=2 to $F$'=3 transition in $^{87}$Rb, which causes atomic fluorescence with the $F$=2 hyperfine-ground-state atoms indicating population. This state-selective detection shows a clear difference in the number of photons scattered with the $F$=1 ground-state atoms. In the second pulse, the light is resonant with the $F$=1 to $F$'=2 repump transition, which rapidly pumps atoms from the $F$=1 hyperfine state up to the $F$=2 level. Between the first and second pulses, the DC bias of the SSB modulator chip is changed dynamically with a single TO phase shifter so that the chip outputs the sideband shifted up in frequency (+1 sideband). For the third pulse, the TO phase shifter is switched back to the -1 sideband output, and the detection beam is again resonant with the $F$=2 to $F$'=3 transition. In the third pulse, all the atoms are in the F=2 state and thus can be used as a normalization signal for the first pulse. In the pulse sequence of state-selective detection, the bias or sideband switching corresponds to switching the quadrature optical phase between π/2 and 3π/2, and the intensity spike at the end of the second pulse is caused by the sudden change in the bias current. This demonstration validates the function of the silicon photonic SSB modulator capable of dynamically modulating the light frequency during a single cycle of a cold-atom system. This state-selective detection is needed to observe the atom interferometer fringes based on atomic population and measure the phase shifts of interest for quantum gravity and inertial sensing [60-61].

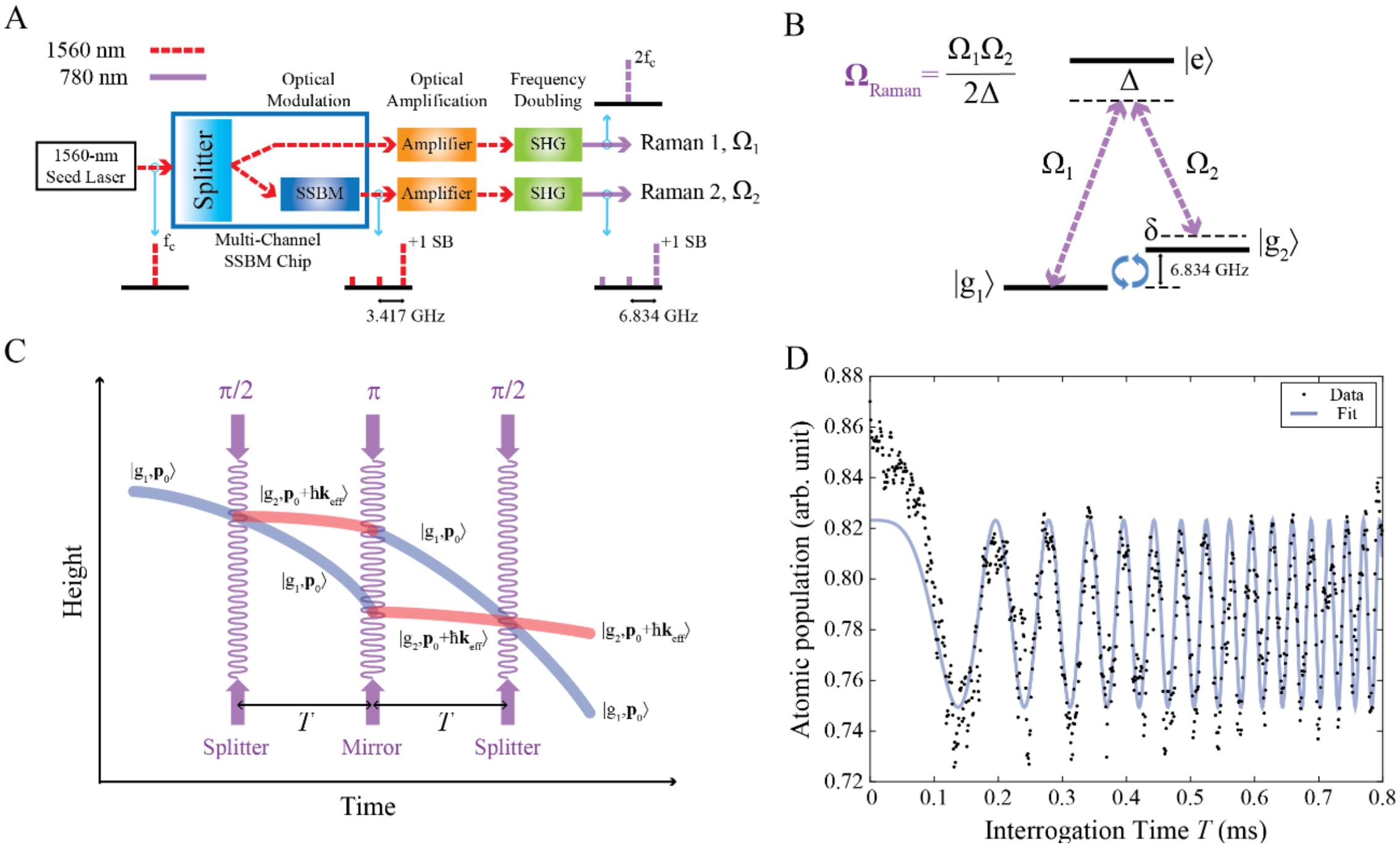

**Fig. 9: Atom interferometry experiment with the use of a multi-channel silicon photonic SC-SSB modulator in an LPAI laser system.** (**A**) Experimental setup for two counter-propagating Doppler-sensitive Raman beams in LPAI operation, where a frequency-doubled optical carrier ($2f_c$) and a frequency-doubled SSB signal (+1 SB) are created at 780 nm with a frequency offset, $f_{HF} \approx 6.834$ GHz (3.417 GHz at 1560 nm). (**B**) Stimulated Raman transitions in a three-level atomic system ($|g_1\rangle$, $|g_2\rangle$, and $|e\rangle$) with an effective Raman Rabi frequency $\Omega_{\text{Raman}}$ where $\Omega_1$ and $\Omega_2$ are single-photon Rabi frequencies for the $|g_1\rangle$-to-$|e\rangle$ and $|g_2\rangle$-to-$|e\rangle$ transition, respectively, and $\Delta$ is a single-photon detuning. A two-photon detuning $\delta$ becomes non-zero when an atomic cloud freely falls under gravitational acceleration. (**C**) Light-pulse sequence, π/2 (splitter) → $T$ → π (mirror) → $T$ → π/2 (splitter) in LPAI operation. Doppler-sensitive Raman beams can coherently address the $|g_1\rangle$ and $|g_2\rangle$, and deliver state-dependent photon recoils between $|g_1, \mathbf{p}_0\rangle$ and $|g_2, \mathbf{p}_0 + \hbar\mathbf{k}_{\text{eff}}\rangle$ for matter-wave interference [60-61], where $\mathbf{p}_0$ is an initial atomic momentum and $\mathbf{k}_{\text{eff}}$ is an effective wavevector related to the photon recoils. (**D**) Atom interferometer fringe produced with the use of a silicon photonic SSB modulator in an LPAI laser system. The data has been smoothed with a four-point running average. The Raman pulse duration was $\tau_\pi = 5$ μs. The estimated local gravitational acceleration is $g \approx 9.77 \pm 0.01$ m/s$^2$, where the error is one standard deviation, as estimated by the nonlinear least-squares fitting routine.

In addition to the single-channel SSB modulator chip, we have also developed a silicon PIC chip (8 mm × 8 mm) that includes four SSB modulators as a multi-channel chip for atom interferometry (see details in section S4). To demonstrate proof-of-concept LPAI operation, we used only one channel of our four-channel modulator chip in an LPAI laser system (Fig. 9A) to produce Doppler-sensitive Raman beams based on a three-level atomic system (Fig. 9B). Two phase-coherent frequency components at 780 nm are generated for rubidium atoms with the use of a single silicon photonic SSB modulator ($f_{HF}/2 \approx 3.417$ GHz) at 1560 nm, which are separated by the ground-state hyperfine splitting ($f_{HF} \approx 6.834$ GHz) at 780 nm after optical amplification and frequency doubling for the two hyperfine ground states, $|\mathrm{g}_1\rangle = |\mathrm{F}=1, \mathrm{m_f}=0\rangle$ and $|\mathrm{g}_2\rangle = |\mathrm{F}=2, \mathrm{m_f}=0\rangle$ (see Fig. 9A-B). Due to the magnetic sub-level $\mathrm{m_f}$ (related to Zeeman splitting) in the ground states, specific light-polarization configuration (e.g., $\sigma^+/\sigma^+$, $\sigma^-/\sigma^-$, or Lin-perp-Lin) is required for the Raman beams.

In the laboratory, the first Raman beam is derived from frequency-doubled carrier light ($2\mathrm{f_c}$ at 780 nm in Fig. 9A) which is offset-locked to the repump transition with -1 GHz detuning, and the second Raman beam is a frequency-doubled SSB light (+1 SB at 780 nm in Fig. 9A) with a frequency offset as $f_{\mathrm{HF}}$. The two Raman beams are combined with crossed linear polarization, and the relative phase of two Raman beams are stabilized by detecting a beat note between the tones and implementing a phase-locked loop that feeds back to the RF source driving the SSB modulator. After the phase lock of the Raman beams, the light-pulse sequence during LPAI operation is generated by an AOM (switching speed ~150 nSec) before delivering the light pulses to atoms. At the sensor head, the two crossed, linearly polarized Raman beams pass through a quarter-wave plate, then they enter the vacuum system and propagate through the cold-atoms. The beams are retro-reflected through a quarter-wave plate to drive counter-propagating, doppler-sensitive Raman transitions. The Raman beam pair is aligned vertically so that the atoms' acceleration due to gravity gives rise to a phase shift which depends on the time interval between pulses in the atom interferometer sequence.

As shown in Fig. 9C, the Raman beams can coherently address the atomic states, $|\mathrm{g}_1\rangle$ and $|\mathrm{g}_2\rangle$, and deliver state-dependent photon recoils along the sensing axis as $|\mathrm{g}_1, \mathbf{p}_0\rangle$ and $|\mathrm{g}_2, \mathbf{p}_0 + \hbar\mathbf{k}_{\mathrm{eff}}\rangle$ for matter-wave interference [60-61], where $\mathbf{p}_0$ is an initial atomic momentum and $\mathbf{k}_{\mathrm{eff}}$ is an effective wavevector related to the photon recoils. An LPAI gravimeter can be realized with a sequence of three light pulses, π/2 (Splitter) → $T$ → π (Mirror) → $T$ → π/2 (Splitter), that drive stimulated Raman transitions to split/redirect/recombine atomic wave-packets for matter-wave interference [60-61], where $T$ is the interrogation time. At the end of the atom interferometer sequence, the relative phase between the two branches of the interferometer is encoded in the atomic population of the two hyperfine-ground states. To observe atom interferometer operation, we measure oscillations in the population of the two ground states as a function the time $T$ between Raman pulses.

In Fig. 9D, we measured the chirped, sinusoidal atom interferometer fringes as the interrogation time $T$ between the Raman pulses ($T$ = 0 to 0.8 ms) is varied. A sinusoidal atom interferometer fringe (resulting from the relative phase change between two Raman beams, *i.e.* the interrogation time scanning) becomes frequency-chirped under gravitational acceleration because an atomic cloud is free-falling in vacuum with respect to two counter-propagating Raman beams delivered from a sensor platform. Therefore, two-photon Raman detuning δ (Fig. 9D) becomes non-zero. As shown in Fig. 9D, the fraction of the atoms in the upper hyperfine state $|\mathrm{g}_2\rangle$ is described by $P_{|\mathrm{g}_2\rangle} \approx P_0 + \frac{c}{2}\cos\left(k_{eff}\, g\, \tau_\pi \left(1+\frac{2}{\pi}\right) T + k_{eff}\, g\, T^2\right)$ where $P_0$ is an offset, $c$ is the interferometer contrast, $g$ is the local acceleration due to gravity, and $\tau_\pi$ is the duration of the mirror pulse [62]. This chirped atom interferometer fringes are obtained with the use of a silicon photonic SSB modulator in an LPAI laser system (Fig. 9B and D). The estimated local gravitational acceleration from the nonlinear least-squares fitting routine is $g \approx 9.77 \pm 0.01$ m/s$^2$, where the error is one standard deviation. To improve the uncertainty and achieve high-performance LPAI operation, we will further investigate silicon photonic SSB modulator at its peak performance (carrier and sideband suppression) along with phase noise in the generated tones and their propagation across various components of the laser system. Moreover, improvement in optical coupling losses (inverse taper couplers), optical packaging, modulator bandwidth, and high speed on-chip optical switching will improve LPAI performance (see supplementary materials). For this demonstration, only a single channel of our four-channel SSB modulator chip was utilized. However, our approach is scalable and simultaneously utilizing all four channels of our four-channel SSB modulator chip for LPAI operation is an active object of future study.

Although much work remains to be done to obtain high-performance atom interferometer with multiple PIC components in a PIC laser architecture, our proof-of-concept demonstration of coherent atomic interference fringes indicates the silicon photonic SSB modulator is a promising advancement toward compact atom-interferometer systems. Furthermore, with the integration of multiple SSB modulators onto a single chip with suppressed carrier and suppressed unwanted sidebands, our silicon photonic SSB modulator chip has the potential to simplify laser systems for atom interferometry and eliminate unwanted systematics. Moreover, with further advances in integration, additional on-chip integration of optical amplifiers, isolators, and frequency doublers will lead to the complete miniaturization of the above atom interferometry laser setup onto a single chip improving both optical losses and component efficiencies, enabling compact and efficient quantum inertial sensors.

## DISCUSSION

We investigated a high-performance silicon photonic SSB modulator and employed this SSB modulator in a PIC laser architecture for quantum sensing. First, we achieved high-performance SSB generation by carefully balancing both optical and RF phases and amplitudes and studied the effects of imbalance on modulator performance. We demonstrated 30 dB carrier suppression and 47.8 dB sideband suppression (over the previous record, 39 dB [44]) near the peak efficiency point of -6.846 dB (20.7%). We note that even higher carrier suppression can be achieved with the use of high-contrast optical splitters. Moreover, we show that the sideband suppression is highly sensitive to the RF amplitude and phase balance between RF channels in a push-pull or dual-drive configured SSB modulator. Because the frequency shift of the SSB modulator is, in principle, variable and limited to the frequency response of the modulator, further refinement using travelling wave electrodes will enable operation of our SSB modulator beyond 20 GHz [63-64]. We also note that the principles of SSB generation outlined here can be applied to different material systems, such as thin-film lithium-niobate, which allows for bandwidths >100 GHz with better linearity than silicon [16,65].

Second, as a proof-of-principle demonstration, we utilized our chip-scale single-channel silicon photonic SSB modulator within an LPAI laser system to demonstrate cold-atom generation in a $^{87}$Rb MOT and state-selective detection. In a rubidium cold-atom system, we showed the capability of our silicon photonic SSB modulator to simultaneously drive both ±1 sidebands for cold-atom generation and perform dynamic optical frequency shifting between the +1 and -1 sidebands for state-selective detection within a single cold-atom cycle.

Lastly, we successfully deployed our fabricated four-channel silicon photonic SSB modulator in an LPAI laser system to demonstrate Doppler-sensitive Raman beams for LPAI operation and measure atom interferometry fringe under gravitational acceleration. Here, we showed an exemplar PIC laser architecture for quantum sensing based on the 1560-to-780 nm frequency doubling approach leveraging hybrid PIC integration (silicon photonics, III-V photonics, and nonlinear photonics) towards a single photonic platform with all the functions necessary for LPAI operation. PIC technologies should enable a complete chip-scale PIC-based laser system with SWaP advantage and improved reliability. However, with the 1560-to-780 nm approach, it is currently challenging to integrate all the PIC components into a single photonic chip due to fabrication and design complexity. The development, maturation, and sharing of modular PIC components will be important to reduce time and cost of PIC development, repurpose them to higher technology-readiness-level applications, increase production volume, and accelerate the realization of a fully-integrated PIC-based laser system for quantum applications through heterogenous and monolithic PIC integration.

## MATERIALS AND METHODS

### Sample fabrication and packaging

The silicon photonic single-sideband modulators were fabricated on 8" silicon-on-insulator (SOI) wafers with 250nm silicon layer on 3um of buried oxide (BOX) at Sandia's Microsystems Engineering, Science and Applications (MESA) complex. The silicon devices were lithographically patterned with a deep ultra-violet (DUV) 248nm ASML laser stepper and subsequently dry-etched. The carrier-depletion modulators were implanted with both arsenic and phosphorous to realize an n-type and p-type doping levels of ~$5x10^{18}/cm^3$ in the p-n junction. The electrical contacts were made to the device using tungsten vias to aluminum top-electrodes [64]. The fabricated devices are subsequently diced into chips then packaged both optically and electrically for testing. Using epoxy, the diced chip is mounted onto a separate interposer silicon chip that consists of gold traces for carrying DC signals. The interposer chip along with the diced chip are then mounted inside a ceramic pin grid array (CPGA) and wire-bonded (automatic) for all DC connections. An RF PCB board with >4 GHz bandwidth is attached to the CPGA. We wire-bond directly from the board to the silicon photonic chip to minimize the RF signal loss. Lastly, for the optical packaging, we carefully align and attach a fiber v-groove array using a UV cured epoxy (Norland Optical Adhesive, NOA 61). See section 4 of supplementary materials for more details.

**Acknowledgments:** We wish to thank Nicholas Boynton for help with packaging, Hayden McGuinness for help with AMO experiments, and Nils Otterstrom for fruitful discussions.
**Funding:** This work is supported by the Laboratory Directed Research and Development program at Sandia National Laboratories, a multi-mission laboratory managed and operated by National Technology & Engineering Solutions of Sandia, LLC, a wholly owned subsidiary of Honeywell International Inc., for the U.S. Department of Energy’s National Nuclear Security Administration under contract DE-NA0003525. This paper describes technical results and analysis. Any subjective views or opinions that might be expressed in the paper do not necessarily represent the view of the U.S. Department of Energy or the United States Government. SAND2022-5279 O.
**Author contributions:** A.L.L, G.B., M.G., C.T.D, and J.L. conceived the research. A.K., G.W.H., and M.G. performed the atom experiments with silicon photonic chip. M.G. designed the silicon photonic chips with input from C.T.D. A.K. and M.G. modelled the single-sideband modulators. A.K and M.G. built the experiment and collected the data for the modulators. A.P., C.D., D.T., A.L.S. contributed to the fabrication of the silicon photonic chips. A.K. and M.G. packaged the photonic chips. G.W.H, J.L., and G.B. investigated atom interferometry integration approach. A.L.L., G.B., J.L., and P.D.D.S supervised the program. A.K., G.W.H., and J.L. wrote the manuscript with input from all authors.

**Competing interests:** Anthony L. Lentine, Grant Biedermann, Michael Gehl, Christopher DeRose, Jongmin Lee, Kevin Michael Fortier granted US Patent (no.: 11545815B1) titled "Compact laser source with frequency modulators generating multiple lines" with the US Patent and Trademark Office (USPTO) on 03 January 2023. These authors declare no other competing interests. All other authors declare they have no competing interests.

**Data and materials availability:** All data needed to evaluate the conclusions in the paper are present in the paper and/or the Supplementary Materials.

# Supplementary Materials for

## High-Performance Silicon Photonic Single-Sideband Modulators for Cold Atom Interferometry

**A. Kodigala *et al.***

Corresponding author: Ashok Kodigala, akodiga@sandia.gov

### 1. Discussion on optical losses

We discuss below the optical losses present in our experimental demonstrations with silicon photonic single-sideband modulator chip. The packaged SiPh modulator chip is operated in conjunction with external erbium-doped fiber amplifiers (EDFAs) and a frequency doubler to produce an optical Raman signal at 780nm for the cold-atom system. The laser system uses a single master/seed laser at 1560 nm as seen in Fig. S1. For demonstration of state-selective detection, the modulator is driven at 1.644 GHz at 1560 nm. The doubled 780nm signal is used to form a magneto-optic trap (MOT). The laser supplies ~0 dBm of optical power followed by an EDFA with 16 dB gain to compensate for the 15 dB insertion loss of the silicon chip. This translates to a coupling loss of ~7.5 dB/facet (input and output) to the silicon chip. There is an additional on-chip loss of ~2-3 dB in the active modulators with -4.7 dB peak conversion efficiency in the single-sideband modulator. This yields approximately -7 dBm of side-band optical power which is further amplified to 20 dBm as input to the doubler. Consequently, there is a few milli-watts of optical power (780nm) at the output of the doubler to probe the atom system.

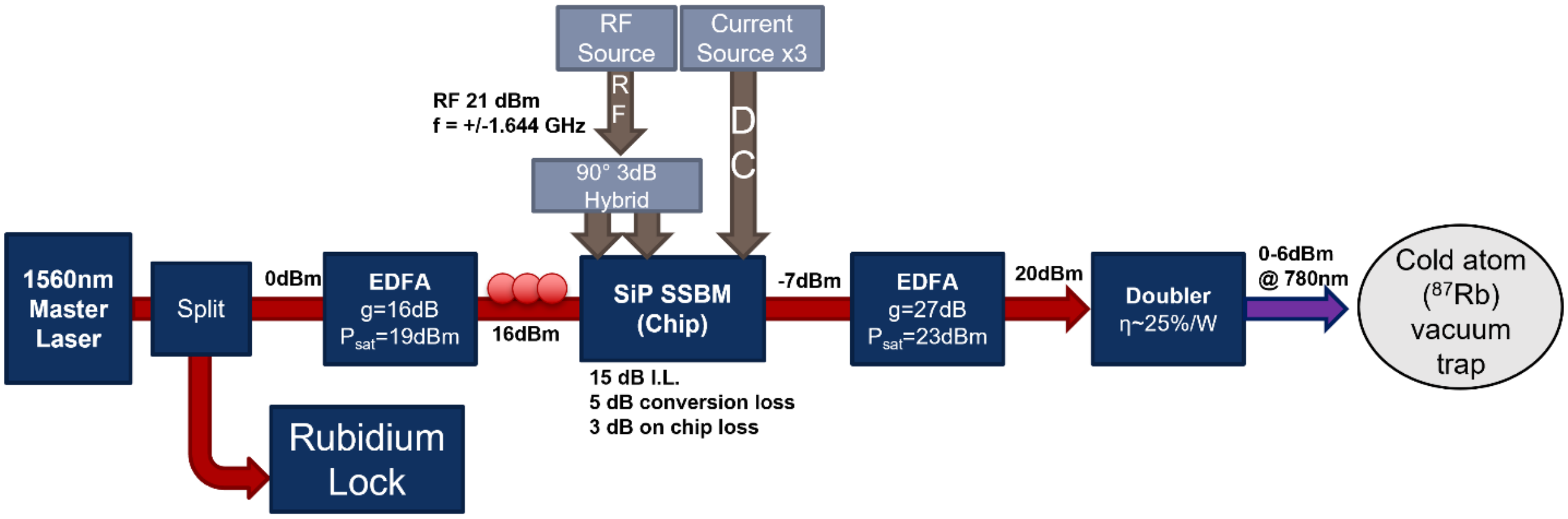

**Fig. S1.** Detailed experimental schematic for state-selective detection with single-sideband (SSB) modulator chips showing all optical losses and conversion efficiencies in line to rubidium atoms. We make use of erbium-doped fiber amplifiers (EDFAs) to compensate for the high optical insertion loss (IL) of the packaged silicon photonic chip.

We have since fabricated devices with both silicon and silicon nitride (SiN) inverse-taper couplers with coupling losses as low as ~2 dB/facet (reduction from the 7.5 dB/facet used in the experiments). With successful packaging of these new devices, we can reduce the amplifiers needed.

### 2. Optical power handling of waveguides

A key consideration of our photonic chip design is the ability to handle the required optical powers. Early on, we determined a requirement for 1mW of optical power in the output fiber of each channel of the modulator. There are many sources of loss from input fiber to output fiber. This includes fiber to chip coupling, propagation loss in the waveguide and loss in the doped regions of the modulator. It also includes intrinsic 'losses', such as the need to split light to 4 channels, and the theoretical maximum conversion efficiency of -4.7 dB. For this reason, relatively high optical power, potentially up to 1 Watt, is required at the input. At high optical powers the transmission of silicon waveguides becomes non-linear. The high optical powers generate two photon absorption (TPA), which increases proportional to the square of the intensity. This in turn creates free carriers which produce free carrier absorption (FCA). To account for this, we conducted high power optical testing of waveguides similar to those used in the modulator. This is shown in Fig. S2(A), where measured output power is plotted as red circles for various input powers up to 400 mW. The data is seen to deviate from linear absorption (dashed black line), and is fit to a model of TPA + FCA (blue curve) in order to extract the relevant material parameters. These parameters are then used to estimate the transmission through our photonic chip as a function of input power and as a function of fiber to chip coupling. This is shown in Fig. S2(B). For poor fiber to chip coupling, very high powers are required to achieve 1 mW output from each channel of the chip. However, since much of the light is lost in coupling, there is no significant non-linear absorption. By reducing fiber to chip coupling, we can dramatically reduce the required input power. At a coupling of 1 dB/facet, we require less than 100mW input to achieve 1mW output (yellow curve). From Figure S2(B), it is clear that at 1mW output power there is not significant non-linear absorption in any of the cases plotted. From the yellow curve, we can see that non-linear absorption does become more apparent when trying to achieve higher output powers, and the output power begins to saturate around 6-7 mW.

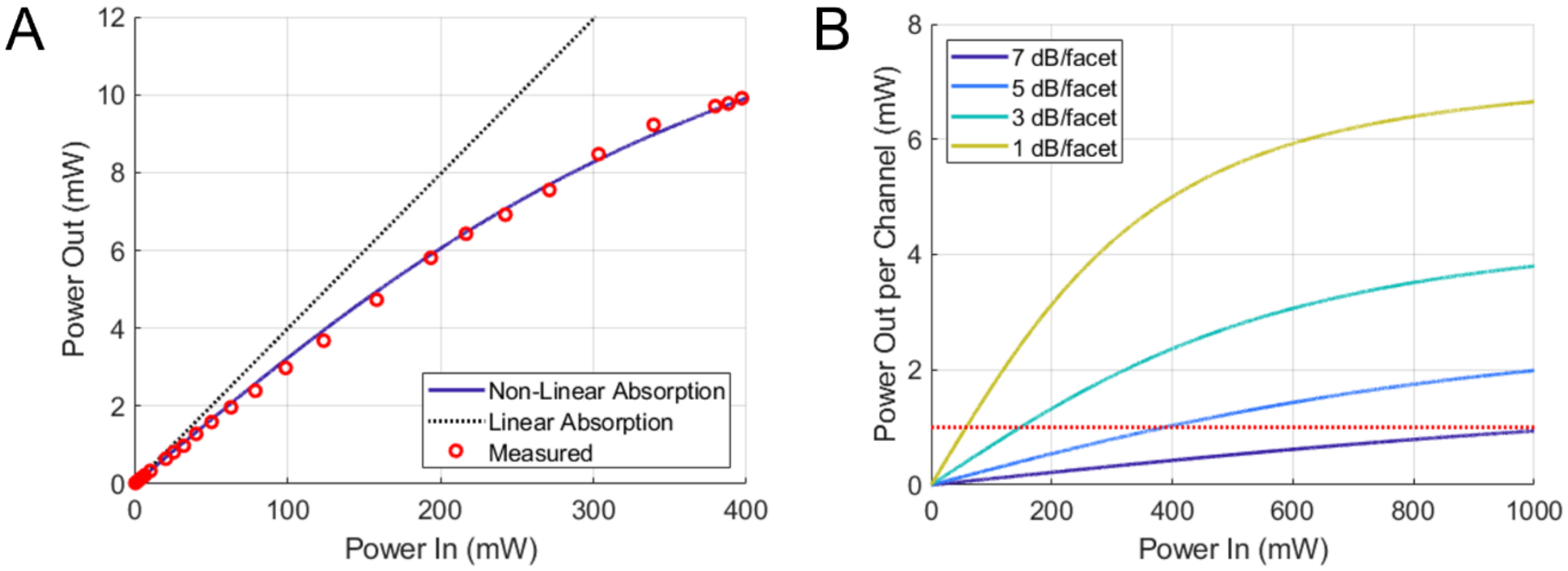

**Fig. S2. Optical power handling of 600nm wide silicon ridge waveguide.** (**A**) Measured output power versus input power of a silicon waveguide and (**B**) simulated output of the photonic chip at high optical inputs for several different fiber to chip coupling efficiencies.

One step taken to improve power handling was the use of silicon nitride when coupling onto the chip. Silicon nitride does not have two photon absorption at a wavelength of 1550nm and can therefore handle higher powers than silicon. Therefore, we couple onto the chip into a silicon nitride waveguide. For a four-channel chip, the power is then split four ways using silicon nitride multi-mode interference splitters before transitioning to silicon. This way the power is already reduced by a factor of 4 before entering silicon. Further increase in optical power handling can be achieved using techniques such as applying a reverse-bias to the waveguides to sweep out free carriers generated by TPA (*66*).

## 3. Discussion on bandwidth of silicon single-sideband modulators

We measure the 3dB optical bandwidth of our silicon Mach-Zehnder modulators used in all the atomic physics experiments to be ~1.13 GHz for a modulator length of 1.55mm (see Fig. S3). These were lumped-element pn-junction modulators. We have since improved the frequency response of these modulators by using travelling-wave segmented electrode designs (*64*). By periodically attaching the capacitive p-n junction to a high impedance travelling-wave electrode, both the impedance and velocity matching can be achieved simultaneously to increase the bandwidth (*67*). We see more than 4 times greater bandwidth from the new design to be ~5 GHz. This frequency response was measured with no reverse DC bias applied to the modulator. We expect even greater bandwidth when a reverse bias is applied, as this increases the depletion region within the doped modulator and reduces capacitance.

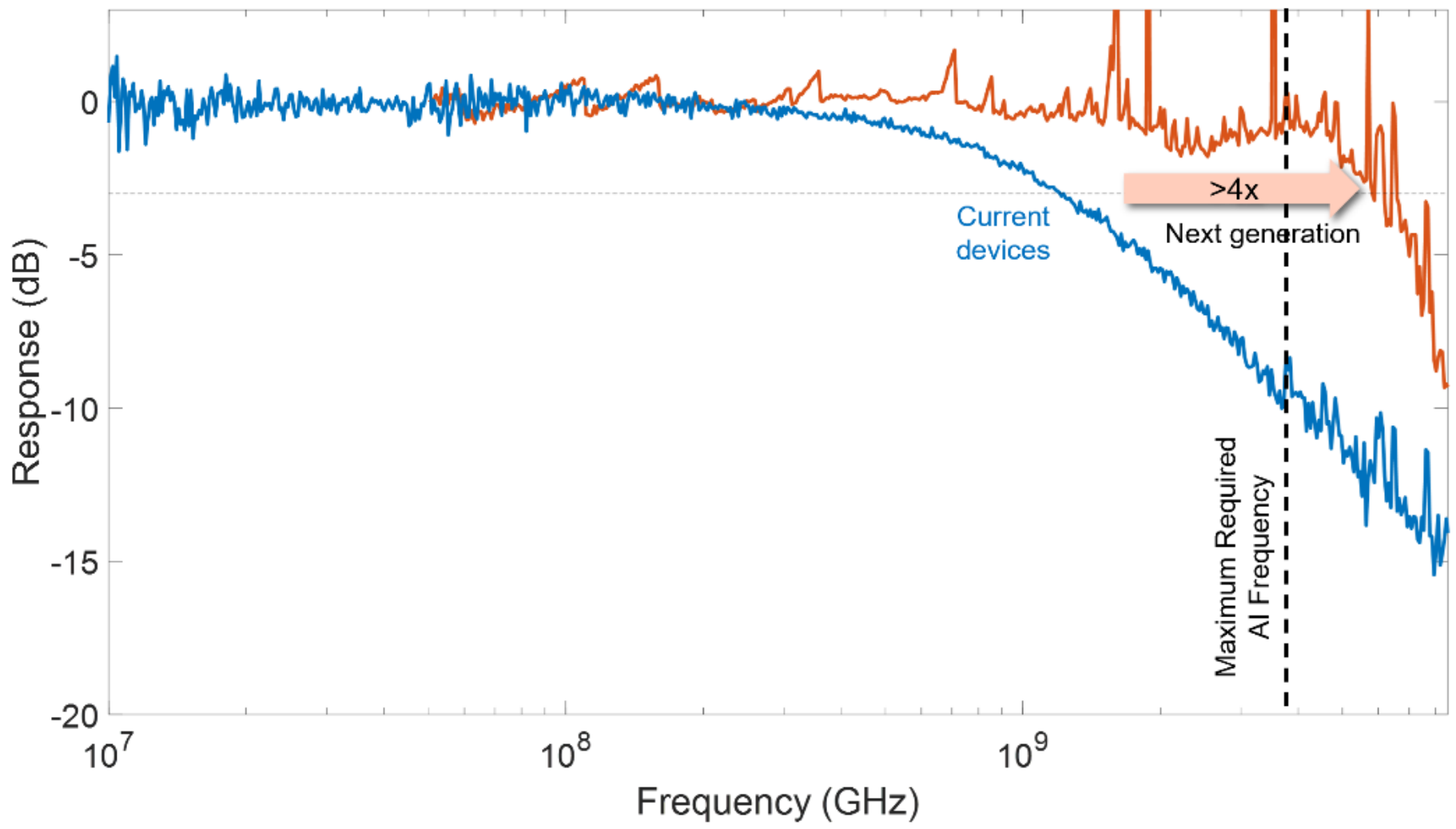

**Fig. S3. Optical frequency response of silicon Mach-Zehnder modulators.** The current modulators have a measured optical 3dB bandwidth of ~1.13 GHz for a modulator length of 1.55mm (blue curve). These packaged devices were used for all the atomic physics experiments. New generation of travelling-wave electrode silicon Mach-Zehnder modulators have an improved optical bandwidth of ~5 GHz at 1560nm (past the ~4 GHz requirement for this atom system) shown as the red curve.

## 4. Packaging of single-sideband modulator chips

To demonstrate practical application of this silicon photonic chip to an atom interferometry system we must package the chip with both electrical and optical interfaces. A cross-section of the packaging geometry is illustrated in Fig. S4. Due to the significant number of electrical I/O required for thermo-optic, electro-optic and photodiode elements in a four-channel modulator chip seen in Fig. S5(A), we implemented a packaging scheme based on a 280-pin count ceramic pin grid array (CPGA). Additionally, this package provided a significantly large die cavity which allowed for placement of both the photonic chip and a fiber array substrate side by side. Direct wire-bonding from the photonic chip to the package would have resulted in very long wire-bonds with potential for shorting and damage. Therefore, an electrical interposer was designed. This was a silicon chip matched to the package die cavity dimensions. The interposer contained a single gold metal layer which routed signals from the package to pads near to the photonic die. Automated wire-bonding was implemented to bond the photonic chip to the interposer and the interposer to the package. Images of a chip after wire-bonding are shown in Fig. S5(B).

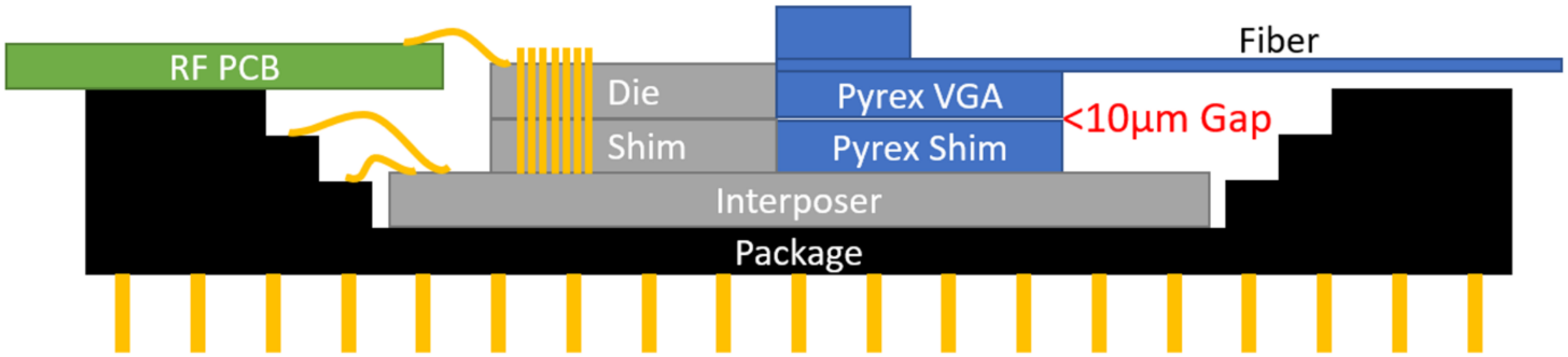

**Fig. S4. Cross section of the packaging scheme implemented.** Packaging incorporates an optical fiber v-groove array (VGA), DC signals (wire-bonds), RF signals from printed-circuit board (PCB), interposer to the silicon photonic die, and ceramic pin grid array (CPGA) package (in black).

While this is suitable for low frequency signals, package and interposer would not be suitable for the high frequency RF signals required to drive the modulator. For that reason, an additional printed circuit board was designed to introduce RF signals. This board featured smp connections, surface mount RF bias tees and co-planar RF waveguides.

This PCB can be seen in Fig. 3 of the main text. The board was characterized by wire-bonding two boards together and measuring RF transmission from one board to the second. It was found to have approximately 4 GHz bandwidth, which is suitable for the frequencies required on this project. The board is attached to the CPGA using a thermally conductive, electrically insulating epoxy. After attachment, wire-bonding is used to route the RF signal from the PCB to the RF pads on the photonic die.

Once a die has been fully packaged electrically, optical packaging is done in a custom packaging setup. For single-channel chips an 8-count fiber array is used for packaging. For four-channel chips a 20-count fiber array is used. As a result of differences in the thickness of the photonic die substrate and the fiber array substrate, additional pieces of silicon and pyrex are used to adjust the heights of each. The silicon photonic die is placed on a silicon shim prior to wire-bonding. An appropriate pyrex shim is then chosen in order to match the height of the fiber array to that of the photonic die as closely as possible. Ideally, when the fiber array is properly aligned to the photonic die, there should be a gap on the order of 10 microns between the fiber array and the shim below it. UV cure epoxy is applied to this gap and subsequently cured. During the curing process the epoxy shrinks by ~1%, causing the fiber array to be pulled downward. We've found that if the layer of epoxy is much greater than 10 microns this shrinkage will cause a significant reduction in optical coupling from fiber to photonic die. Additional epoxy is applied to the interface between the photonic die and fiber array. This provides an index matching effect which reduces loss at this interface and provide additional mechanical support. Lastly, UV cure epoxy is used to attach the fibers to the edge of the ceramic package for additional strain relief.

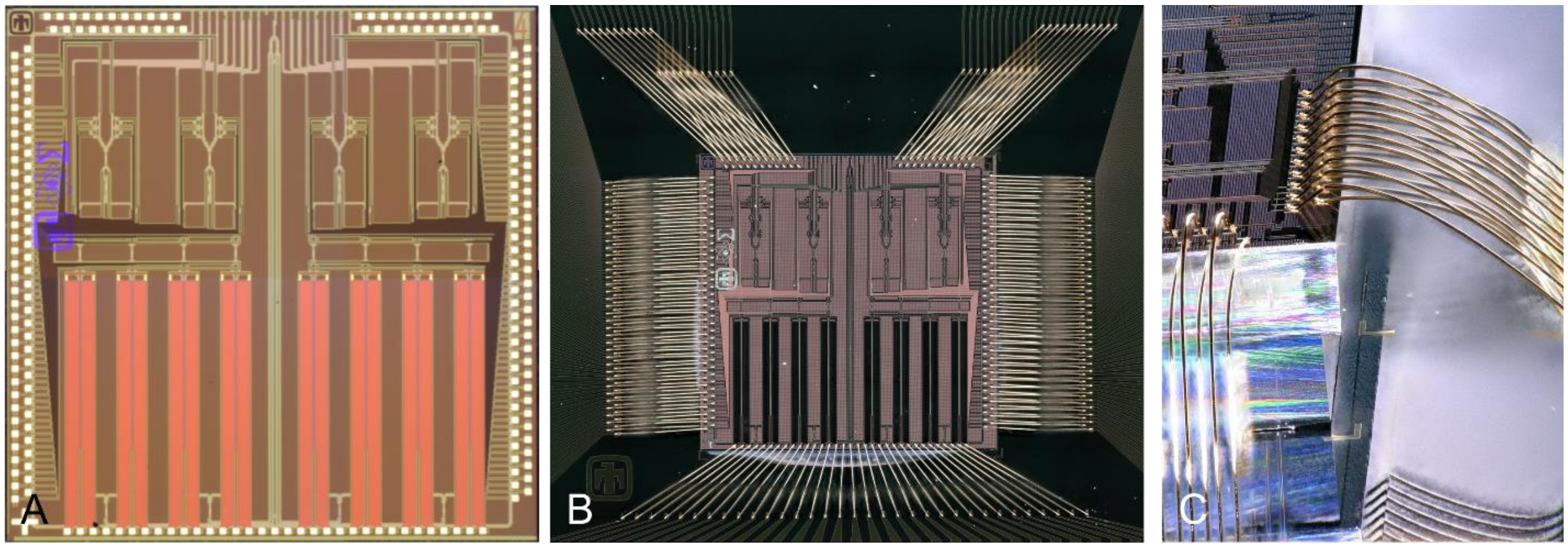

**Fig. S5. Four-channel single-sideband modulator chip.** (**A**) Micrograph image of fabricated 4-channel single-sideband modulator chip with 8mm-by-8mm footprint. (**B**) Top view of a four-channel silicon photonic die following automated wire-bonding. (**C**) Side view of the same die showing the silicon shim below the die.

During alignment of the fiber array, an alignment waveguide loop on the chip is used to monitor fiber to chip coupling. This alignment waveguide connects the first and last fiber in the array, such that a laser can be transmitted through the chip and monitored for alignment and curing. It was noted that the transmission through this alignment loop was approximately -17dB prior to curing of the epoxy. The epoxy was cured overnight, and the transmission was observed to improve to approximately -16dB. Propagation of light within the alignment loop contributes approximately 1dB of loss, so we estimate a fiber to chip coupling of -7.5 dB.